\documentclass[aps,prb,reprint,twocolumn,superscriptaddress,showpacs]{revtex4-1}
\usepackage[T1]{fontenc}
\usepackage[utf8]{inputenc}
\usepackage{amsmath}
\usepackage{amssymb}
\usepackage{braket}
\usepackage{graphicx}
\usepackage{hyperref}
\usepackage{color}
\usepackage{multirow, makecell}
\usepackage{subcaption}

\newcommand{\abs}[1]{\left| #1 \right|}
\renewcommand{\vec}[1]{\boldsymbol{#1}}

\DeclareMathOperator{\Int}{Int}

\begin{document}

\title{Quasiparticle poisoning effects on the dynamics of topological Josephson junctions}

\author{Daniel Frombach}
\affiliation{Institut f\"ur Mathematische Physik, Technische Universit\"at Braunschweig, D-38106 Braunschweig, Germany}

\author{Patrik Recher}
\affiliation{Institut f\"ur Mathematische Physik, Technische Universit\"at Braunschweig, D-38106 Braunschweig, Germany}
\affiliation{Laboratory for Emerging Nanometrology Braunschweig, D-38106 Braunschweig, Germany}

\begin{abstract}
The fractional Josephson effect remains one of the decisive hallmarks of topologically protected Majorana zero modes. We analyze the effects of parity violating quasiparticle poisoning onto the current voltage characteristics of topological Josephson junctions. We include poisoning events directly within the resistively shunted junction (RSJ) model in the overdamped limit both in the short- and long-junction regime. We calculate the current voltage characteristics numerically where poisoning is modeled either via additional rates in the Fokker-Planck equations or by a time dependent parity and compare them to the limits of no and strong poisoning rates which we obtain analytically. Combining the tilted washboard potential with poisoning events, we show that the critical current of the long junction limit can be used as a probe of the junction topology even in the high temperature poisoning case where relaxation- and excitation processes are equally likely. Using the tilted washboard potential model we develop three different schemes to measure the poisoning rate thereby also extending the consideration to two pairs of helical edge states containing a constriction that allows for tunneling between the two pairs of edge states.
\end{abstract}

\date{\today}
\maketitle

\section{Introduction}
\label{sec:introduction}

Majorana zero modes \cite{Kitaev2001, Qi11, Alicea12, Mourik12, Das12, BeenakkerRev13} have been proposed as the basis for intrinsically fault tolerant topological qubits \cite{Nayak2008, Alicea11, Hoffman2016, PengParAn16, Karzig2017, Plugge17}. One of their key signatures is the fractional Josephson effect. In contrast to topologically trivial Josephson junctions, which typically feature a $ 2\pi $ periodic current phase relation, topological Josephson junctions exhibit a $ 4\pi $ periodicity \footnote{We consider the non interacting case. When interactions are considered, $Z_4$ parafermions \cite{Zhang14, Orth15, Klinovaja2015, Alicea16, VinklerAviv17, Fleckenstein19} can form instead of Majorana zero modes resulting in an $ 8\pi $ periodicity \cite{Zhang14}.} in its current phase relation \cite{Kitaev2001, Kwon04, Fu2009, Alicea11, Rokhinson2012, Kane15, Laroche2017}.
While such a $ 4\pi $ periodicity can also arise from other effects \cite{Kwon04, Michelsen08, Zazunov18, Chiu19}, it is one signature hinting at the existence of Majorana excitations.
Furthermore, experimental observation of the fractional Josephson effect is complicated by the coupling of the Josephson junction to the environment. The $ 4\pi $ periodicity in the fractional Josephson effect is protected by fermion parity conservation \cite{Fu2009}, that is, as long as the fermion parity of the junction stays fixed, a topological nontrivial Josephson junction will exhibit a $ 4\pi $ periodicity. However, in realistic settings electrons can tunnel into (out of) the junction from (into) external electron reservoirs \cite{Flensberg2010, Budich12}, an effect known as quasiparticle poisoning \cite{Maennik04, Aumentado04, Rainis2012, Leijnse12}. This process changes the fermion parity of the junction and fermion parity conservation is broken. As a result, the $ 4\pi $ periodicity is no longer protected and will generally break down to a $ 2\pi $ periodicity.

Despite this, elaborate experimental schemes have been theoretically proposed \cite{Badiane11, Heck11, SanJose12, Dominguez12, Houzet13, SanJose13, Sau17} and first signatures of the fractional Josephson effect have been reported \cite{Rokhinson2012, Wiedenmann2015, Bocquillon16, Laroche2017, Deacon17, LeCalvez2019}. To overcome the problem of quasiparticle poisoning these setups focus on dynamical properties of fractional Josephson junctions. For instance, if the driving current of a normal Josephson junction contains a time periodic part, voltage plateaus known as Shapiro steps will emerge in the current voltage characteristic at integer multiples of a base voltage $ V_n = n \hbar \omega / 2e $ given by the frequency $ \omega $ of the applied driving current \cite{Shapiro63, Russer1972, bookTinkham1996}. In the case of topological junctions the $ 4\pi $ periodicity manifests itself in the absence of odd steps \cite{Kitaev2001, Rokhinson2012}. Another proposal \cite{Beenakker2013} instead uses the fact, that in a static dc current measurement the critical current of a topological Josephson junction is predicted to be twice as large as the critical current of a topologically trivial junction provided that the temperature of the poisoning particles is low and the junction is in the long junction regime, that is a junction where the distance between the two superconductors $ L $ is much larger than the superconducting coherence length $ \xi_{sc} = \hbar v_F / \Delta $ with $ v_F $ being the Fermi velocity of the helical edge states \cite{Koenig07, Knez12, Hart2014} mediating the junction and $ \Delta $ being the superconducting gap. Although much research has been done on short junctions \cite{Lutchyn10, Pikulin12, Virtanen13, Cayao2017}, long junctions are predicted to show interesting features \cite{Beenakker2013, Zhang14, Crepin14}. Despite that, to the best of our knowledge, a detailed investigation of the dynamics of such long fractional Josephson junctions in the presence of quasiparticle poisoning has not been performed so far,
and it is one of the main goals of this article to look into this task.

In this manuscript, we analyze the effects of quasiparticle poisoning on the above mentioned experimental schemes via two different methods based on a resistively shunted junction (RSJ) model \cite{bookTinkham1996, Dominguez17, PicoCortes17, Feng18, LeCalvez2019}. First, we rewrite the RSJ model in terms of Fokker-Planck equations \cite{Lee2014, Peng2016}. Because these have the form of a master equation, we can include parity changing terms into them by adding additional terms corresponding to poisoning rates (PR) that describe quasiparticles tunneling in or out of the system thereby changing the particle number parity. Following Ref.~ \onlinecite{Lee2014}, we model these rates by a coupling constant $\Gamma$ multiplied by a Fermi distribution characterized by a poisoning temperature $T_b$ (see Eq.~(\ref{eq:poisoning_rate})).
$\Gamma$ will be compared to the inverse of the characteristic time scale of the junction $\tau_J$ defined in Sec.~\ref{sec:model} whereas $k_BT_b$ will be compared to the characteristic energy scale associated with the Andreev levels in the junction, which is either $\Delta$ (short junction) or the Thouless energy $\hbar v_F/L$. Accordingly, we refer in the article to weak/strong PRs and low/high poisoning temperatures (PT). Note that the PT can be different from the junction temperature $T$ \cite{Lee2014} and we treat the two as independent parameters.

The Fokker-Planck approach in the presence of poisoning events is suitable at finite junction temperature $T$. Another approach will be to assume a time dependent parity of the junction and to numerically solve the resulting time dependent RSJ model, which essentially simulates the junction at zero temperature. Both methods will be described in detail in Sec.~\ref{sec:model}, followed by an analytical solution of the Fokker Planck equations for effective models in the large PR regime and a numerical study for the intermediate PR regime in Sec.~\ref{sec:poisoningEffect}. In Sec.~\ref{sec:ciritcalCurrents}, we investigate the effects of poisoning events onto the critical currents of different Josephson junctions and devise schemes to measure the PR in Sec.~\ref{sec:poisoningRates} using the method of time-dependent parities. An extended Josephson junction where a magnetic flux can be threaded between the two superconductors is discussed in Sec.~\ref{sec:extendedJunction}. In the presence of a constriction, the parity of each helical edge state can be changed by a tunneling event whereas the total parity stays constant. We identify a tunneling resonance for small PR situated at exactly half the critical current with a voltage peak having a height that is determined by the tunneling rate and a width that is bounded by the intrinsic PR of each edge. Finally, we summarize and discuss our results in Sec.~\ref{sec:conclusion}.

\section{Model}
\label{sec:model}

\begin{figure}
\includegraphics[width = \columnwidth]{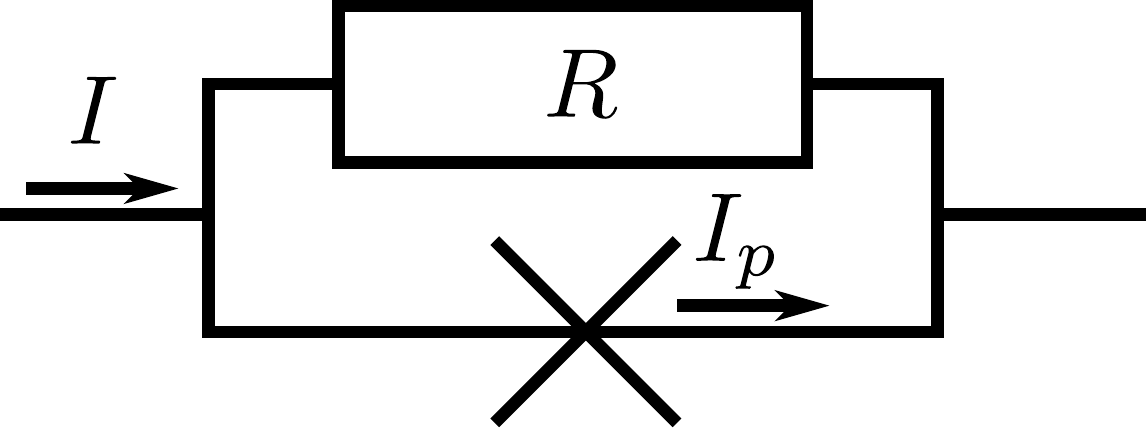}
\caption{The Josephson junction is modeled by an ideal Josephson junction (cross) which has been resistively shunted (resistor with resistance $R$). The junction is current biased by the current $ I $. The current through the entire circuit is the sum of the current through the ideal Josephson junction ($ I_p $) and the current through the resistor $I-I_p$.}
\label{fig:RSJ}
 \end{figure}

We model a Josephson junction embedded in an electromagnetic environment with the resistively shunted junction (RSJ) model \cite{bookTinkham1996}, where the physical Josephson junction is described by an ideal Josephson junction connected in parallel with a resistance $ R $ into a circuit (Fig.~\ref{fig:RSJ}). The existence of a geometric capacitance $ C $ of the junction can be neglected if the time scale of the resulting $ RC $-circuit $ \tau_{RC} = RC $ is small compared to the intrinsic time scale of the junction $ \tau_J = \frac{\hbar}{2e} \frac{1}{I_c R} $ with $ I_c $ being the critical current of the junction \cite{bookTinkham1996, Beenakker2013}. The explicit values for $ R $ and $ C $ depend on the specific experimental setup, however typical values range between $ R \sim 50 - 150 \, \Omega $ and $ C \sim 1 $ fF $ - \, 1 $ aF \cite{bookTinkham1996, Oostinga2013, Wiedenmann2015}. The junction in Ref.~\citenum{Oostinga2013} falls into this category as well as the junction from Ref.~\citenum{Wiedenmann2015} assuming the geometric capacitance $ C $ is smaller than $ \sim 10^{-14} $ F. 

The current through the circuit is given by Kirchhoff's rule \cite{bookTinkham1996}
\begin{equation}
I = I_p(\phi) + \frac{V}{R},
\end{equation}
where $ I_p(\phi) $ is the current through the ideal Josephson junction under consideration, where $ p = 0, 1 $ denotes the parity of the junction. The second term describes the current flowing through the resistor with resistance $ R $. Depending on the driving current $ I $ a voltage $ V $ can develop across the junction. Using the Josephson relation $ V = \hbar\dot{\phi}/(2e) $ the RSJ model can be rewritten as
\begin{equation}
\label{eq:model:RSJDGL}
I = I_p(\phi) + \frac{\hbar}{2eR} \dot{\phi}
.
\end{equation}
The above description holds for a specific conserved parity $ p $ where $p=0$ (even) and $p=1$ (odd). We can directly include parity fluctuations due to quasiparticle poisoning in the RSJ model. We investigate this route using two methods which we discuss in the following.

\subsection{Fokker-Planck equation}

We rewrite Eq.~\eqref{eq:model:RSJDGL} using the relation $ I_p(\phi) = \frac{e}{\hbar} \partial_\phi E_p(\phi) $ with the Josephson energy $ E_p(\phi) $ as

\begin{equation}
\label{eq:model:fokkerPlanck:dgl}
\dot{\phi} = - \frac{\mu}{2} \partial_\phi U_p + \mu \zeta(t)
\end{equation}

with $ \mu = \frac{4e^2R}{\hbar^2} $ and the washboard potential \cite{bookTinkham1996}

\begin{equation}
U_p = E_p(\phi) - \frac{\hbar I}{e} \phi
.
\end{equation}

Here, we have also included a random thermal driving $ \zeta(t) $ at temperature $ T $ which is uncorrelated in time \cite{Lee2014, Peng2016}

\begin{equation}
\langle \zeta(t) \zeta(t') \rangle = \frac{k_B T \hbar^2}{2 R e^2} \delta(t - t')
.
\end{equation}

Solving the differential equation \eqref{eq:model:fokkerPlanck:dgl} is equivalent to solving the Fokker-Planck equation \cite{Ambegaokar1969, Haenggi1990}

\begin{equation}
\label{eq:model:fokkerPlanck:FPEnoPoisoning}
\partial_t \mathcal{P}_p = \mu \partial_\phi \left[
\left( \frac{1}{2} \partial_\phi U_p \right) \mathcal{P}_p
+ k_B T \partial_\phi \mathcal{P}_p
\right],
\end{equation}

where $ \mathcal{P}_p $ is the probability density of finding the junction at phase $ \phi $ in parity $ p $ and $ T $ is the junction temperature. Both parity sectors are described by a single Fokker-Planck equation and are completely independent of each other. The Fokker-Planck equation~\eqref{eq:model:fokkerPlanck:FPEnoPoisoning} has the form of a master equation. We can therefore model poisoning effects by adding additional terms in the form of rates on the right hand side \cite{Lee2014}
\begin{equation}
\begin{aligned}
\label{eq:model:fokkerPlanck:FPE}
\partial_t \mathcal{P}_p = \mu \partial_\phi &\left[
\left( \frac{1}{2} \partial_\phi U_p \right) \mathcal{P}_p
+ k_B T \partial_\phi \mathcal{P}_p
\right] \\
& + \sum_{p'} W_{p'p} \mathcal{P}_{p'} 
- W_{pp'} \mathcal{P}_p,
\end{aligned}
\end{equation}
where
\begin{equation}
\label{eq:poisoning_rate}
W_{pp'} = \Gamma f\left[ \frac{U_{p'} - U_p}{k_B T_b} \right]
\end{equation}
is the rate at which the probability density $ \mathcal{P}_p $ passes over into the probability density $ \mathcal{P}_{p'} $ \footnote{
Due to multiple Andreev bound states inside of long junctions multiple configurations are possible within a given parity sector.
We assume the relaxation time within a given parity sector to be faster than all other time scales in the system \cite{Olivares2014}, so that the junction evolves according to the ground state for a given parity sector at junction temperatures $ k_B T < E_T $.
}.
Here $ \Gamma $ is a constant poisoning rate, $ f[x] $ the Fermi function and $ T_b $ is the temperature of the poisoning electrons which we consider to be an independent parameter.
The voltage that develops across the junction is now given by
\begin{equation}
\label{eq:model:fokkerPlanck:voltage}
\begin{aligned}
V &= \frac{\hbar}{2e} \langle \dot{\phi} \rangle
= \frac{\hbar}{2e} \sum_p \int_0^{4\pi} d\phi \,  \dot{\phi} \mathcal{P}_p \\
&= -\frac{eR}{\hbar} \sum_p \int_0^{4\pi} d\phi \, (\partial_\phi U_p) \mathcal{P}_p.
\end{aligned}
\end{equation}

\subsection{Time dependent parity}
\label{subsec:timeDependentParity}

The above method assumes a finite junction temperature $ T $. State of the art experiments can however reach temperatures far below the superconducting gap $ \Delta $ as well as far below the Thouless energy $ E_T = \hbar v_F / L $ \cite{Oostinga2013, Wiedenmann2015} where $ v_F $ is the Fermi velocity of the host material and $ L $ the distance between the two superconductors forming the Josephson junction. It will therefore be interesting to look at the zero temperature limit. However, in this limit the Fokker-Planck equations~\eqref{eq:model:fokkerPlanck:FPE} no longer hold for currents below the critical current so that another method needs to be developed in order to analyze this regime\footnote{
In order for the Fokker-Planck equations to hold both $ \partial_{\phi} U_p $ and $ \mathcal{P}_p $ need to be almost constant on the thermal length scale $ \sqrt{\frac{k_B T}{\xi} \frac{RC}{\tau_J}}$ \cite{Haenggi1990}.
For driving currents below the critical current the probability density $ \mathcal{P}_p $ develops a divergence for vanishing temperature $ T $ such that this requirement no longer holds.
}.

To this end we include poisoning effects into the RSJ model by assuming a time dependent parity $ p(t) $.
The parity takes the discrete values $ 0 $ or $ 1 $ and switches between the two at specific times $ t_i $ with the probability $ f\left[ \frac{U_{p'} - U_p}{k_B T_b} \right] $.
The switch is assumed to be instantaneous, i.e. the time scale over which a single poisoning electron tunnels into or out of the junction is much shorter than the intrinsic time scale of the junction.
The $ n = \Int(\Gamma \tau) $
\footnote{Here $ \Int(\cdot) $ is rounding to the nearest integer.}
times $ t_i $ are randomly (uniform distribution) selected out of a time interval $ [0, \tau] $.
Inserting this time dependent parity $ p(t) $ into and numerically integrating the differential Eq.~\eqref{eq:model:RSJDGL} allows to calculate the voltage
\begin{equation}
V = \frac{\hbar}{2e} \langle \dot{\phi} \rangle
= \frac{\hbar}{2e} \frac{1}{\tau} \int_0^{\tau} dt \, \dot{\phi}
= \frac{\hbar}{2e} \frac{\phi(\tau) - \phi(0)}{\tau}
\end{equation}
developing across the junction ($ \tau \gg \tau_J; \Gamma^{-1} $).

\subsection{Parameters of a single topologically non-trivial junction}

The energy phase relation for the short junction limit is given by \cite{Fu2009}
\begin{equation}
E_p(\phi) = (-1)^p \Delta \cos \left (\frac{\phi}{2} \right)
\end{equation}
and in the long junction limit by $(\phi \in [0,4\pi) \mod 4\pi)$
\begin{equation}
\label{eq:model:longJunctionEPR}
E_p(\phi) = \frac{E_T}{4\pi} \left\{ \begin{aligned}
&\left\{ \begin{aligned}
&\phi^2 && \phi \in [0, 2\pi) \\
&(\phi - 4\pi)^2 && \phi \in [2\pi, 4\pi) \\
\end{aligned} \right. && p = 0 \\
&(\phi - 2\pi)^2 && p = 1
\end{aligned} \right.
\end{equation}
which can be derived by integrating the current phase relation in the limit $ T \rightarrow 0, L \rightarrow \infty $ given in Ref.~\citenum{Beenakker2013}.
Eq.~\eqref{eq:model:longJunctionEPR} also holds to a good approximation for finite temperatures if $ k_B T < E_T $.
From here on the current voltage characteristic will be given in terms of the normalized voltage $ \nu = V / (I_c R) $ and current $ x = I / I_c $ where $ I_c = (e / \hbar) \xi $ and $ \xi = \Delta \, (E_T) $ in the short (long) junction limit.
Furthermore, the PR $ \Gamma $ can be expressed in terms of the rescaled dimensionless PR $ \gamma = \Gamma \tau_J $ where $ \tau_J = 2 / (\mu \xi) $ is the intrinsic time scale of the junction \cite{Russer1972}.
Intrinsic quasiparticle poisoning events take place on the time scale $ \tau_{qp} \sim 1 \, \mu  $s \cite{Rainis2012} whereas the intrinsic timescale of the junction $ \tau_J $ is on the order of $ \tau_J \sim 10 $ ps \cite{Beenakker2013, Oostinga2013, Wiedenmann2015} so that intrinsic quasiparticle poisoning should always be slow compared to the intrinsic time scale of the junction, i.e. $ \gamma \ll 1 $.
However, it will still be important to distinguish intermediate and fast poisoning regimes, because artificially introduced tunable parity conservation breaking sources can potentially achieve larger PRs \cite{Frombach18}.
Furthermore, we will later introduce a scheme to artificially decrease the intrinsic time scale of the junction $ \tau_J $ (see Fig.~\ref{fig:RSJRExt}).

\section{Poisoning Effects including thermal noise}
\label{sec:poisoningEffect}

First, we investigate the effect that poisoning has on the current voltage characteristics analytically in the limit of large PRs without thermal noise, and numerically for finite junction temperature $ T $ by numerically solving the Fokker-Planck Eqs.~\eqref{eq:model:fokkerPlanck:FPE}. 

\subsection{Analytical limit for strong poisoning rates and zero junction temperature}

\begin{table}
	
\begin{center}
	
\renewcommand{\arraystretch}{2}
		
\begin{tabular}{ | c | c | c | }
\hline
Junction Type & Poisoning & Voltage \\
\hline
\multirow{3}{*}[-0.2cm]{Short}
& no & 
\multirow{2}{*}{$ \nu = \sqrt{x^2 - \left( \frac{1}{2} \right)^2} $} \\
\cline{2-2}
& low temp. & \\
\cline{2-3}
& high temp. & 
\vphantom{$ \sqrt{x^2 - \left( \frac{1}{2} \right)^2} $} 
$ \nu = x $ \\
\hline
\multirow{3}{*}[-0.2cm]{Long}
& no & 
$ \nu = 2 \left[ 
\ln \left( \frac{x + 1}{x - 1} \right)
\right]^{-1} $ \\
\cline{2-3}
& low temp. & 
\multirow{2}{*}{ $ \nu = \left[ 
\ln \left( \frac{x + \frac{1}{2}}{x - \frac{1}{2}} \right)
\right]^{-1} $} \\
\cline{2-2}
& high temp. & \\
\hline
\end{tabular}
		
\end{center}

\caption{Analytical solution of the current voltage characteristics for different models of effective potentials (see text) in the limit of vanishing junction temperature $ T \rightarrow 0 $. For low PTs the effective potential $ U_{\textrm{eff}} = \min_p \left[ U_p \right] $ was used whereas for high PTs the effective potential $ U_{\textrm{eff}} = \frac{U_0 + U_1}{2} $ was used.}
\label{tab:poisoningEffect:solutions}

\end{table}

In the limit of vanishing temperature $ T \rightarrow 0 $ analytical solutions (Tab~\ref{tab:poisoningEffect:solutions}) can be obtained by treating the poisoning, if present, via an effective potential $U_{\textrm{eff}}$, i.e. by dropping the index $ p $ and setting $ W_{pp'} \rightarrow 0 $ and $ U_p \rightarrow U_\textrm{eff} $ in Eq.~\eqref{eq:model:fokkerPlanck:FPE}\footnote{
These analytical limits can also be obtained from the original differential equation \eqref{eq:model:RSJDGL} describing the dynamics of the RSJ model.
}. For this we follow arguments outlined in Ref.~\citenum{Lee2014} and extend them also to the long junction limit. For vanishing PRs, this effective potential is simply given by the parity constrained potential $ U_p $ depending on the fixed parity $p$ of the junction. For large PRs, however, two cases can be distinguished. In the case where the poisoning temperature (PT) $ T_b $ is low compared to the intrinsic energy scale of the junction $ \xi $ the poisoning events will preferentially relax the system to its instantaneous ground state so that if the poisoning events are frequent enough, i.e. if $ \Gamma $ is large compared to the intrinsic time scale of the junction, the system will effectively follow the potential \cite{Lee2014}
\begin{equation}
U_{\textrm{eff}} = \min_p \left[ U_p \right]
.
\end{equation}
In the case of high PT ($ T_b \gg \xi $) all poisoning events can also excite the system with equal probability as compared to relaxation events so that if the poisoning events are frequent enough, the system will effectively follow the potential
\begin{equation}
U_{\textrm{eff}} = \frac{U_0 + U_1}{2}
.
\end{equation}

\subsection{Finite junction temperature}

The current voltage relation at finite junction temperature $T$ can be obtained by numerically solving the Fokker-Planck Eqs.~\eqref{eq:model:fokkerPlanck:FPE} and evaluating Eq.~\eqref{eq:model:fokkerPlanck:voltage}.

\begin{figure*}
	\begin{subfigure}{\columnwidth}
		\includegraphics[width = \columnwidth]{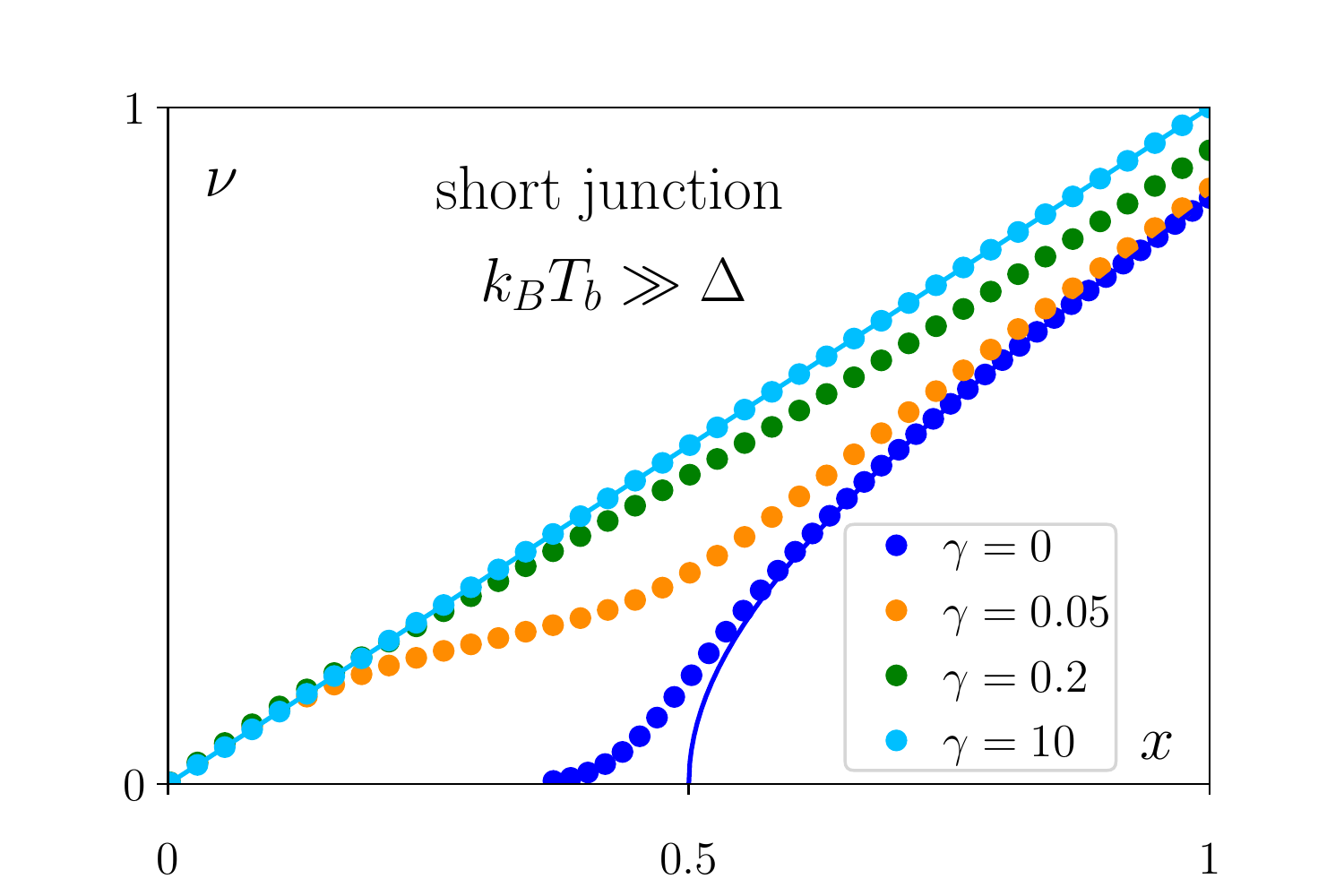}
		\label{fig:poisoningPaths:shortIVFPHighTemp}
	\end{subfigure}
	\hspace{-0.9cm}%
	\begin{subfigure}{\columnwidth}
		\includegraphics[width = \columnwidth]{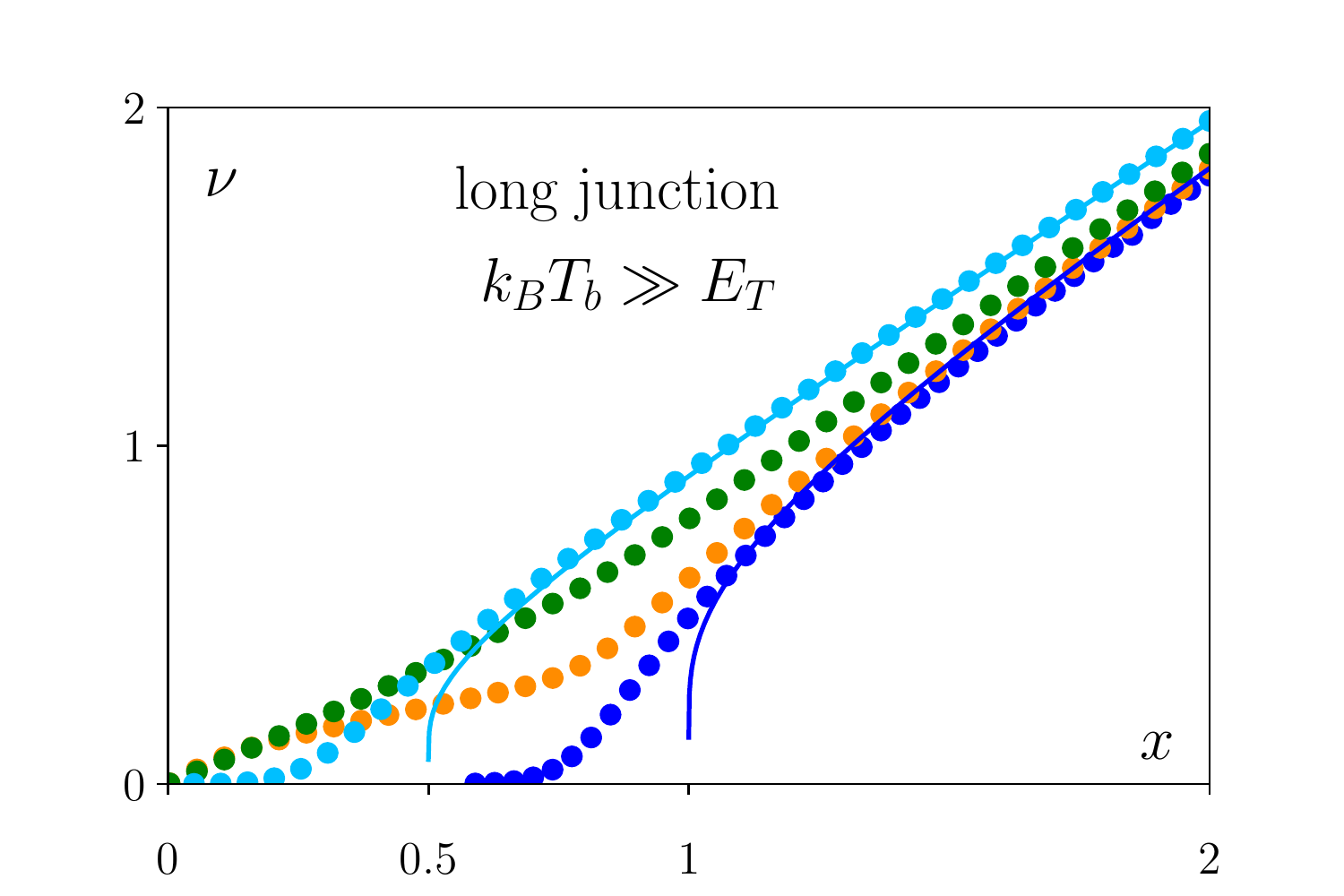}
		\label{fig:poisoningPaths:longIVFPHighTemp}
	\end{subfigure}
	\begin{subfigure}{\columnwidth}
		\vspace{-0.5cm}%
		\includegraphics[width = \columnwidth]{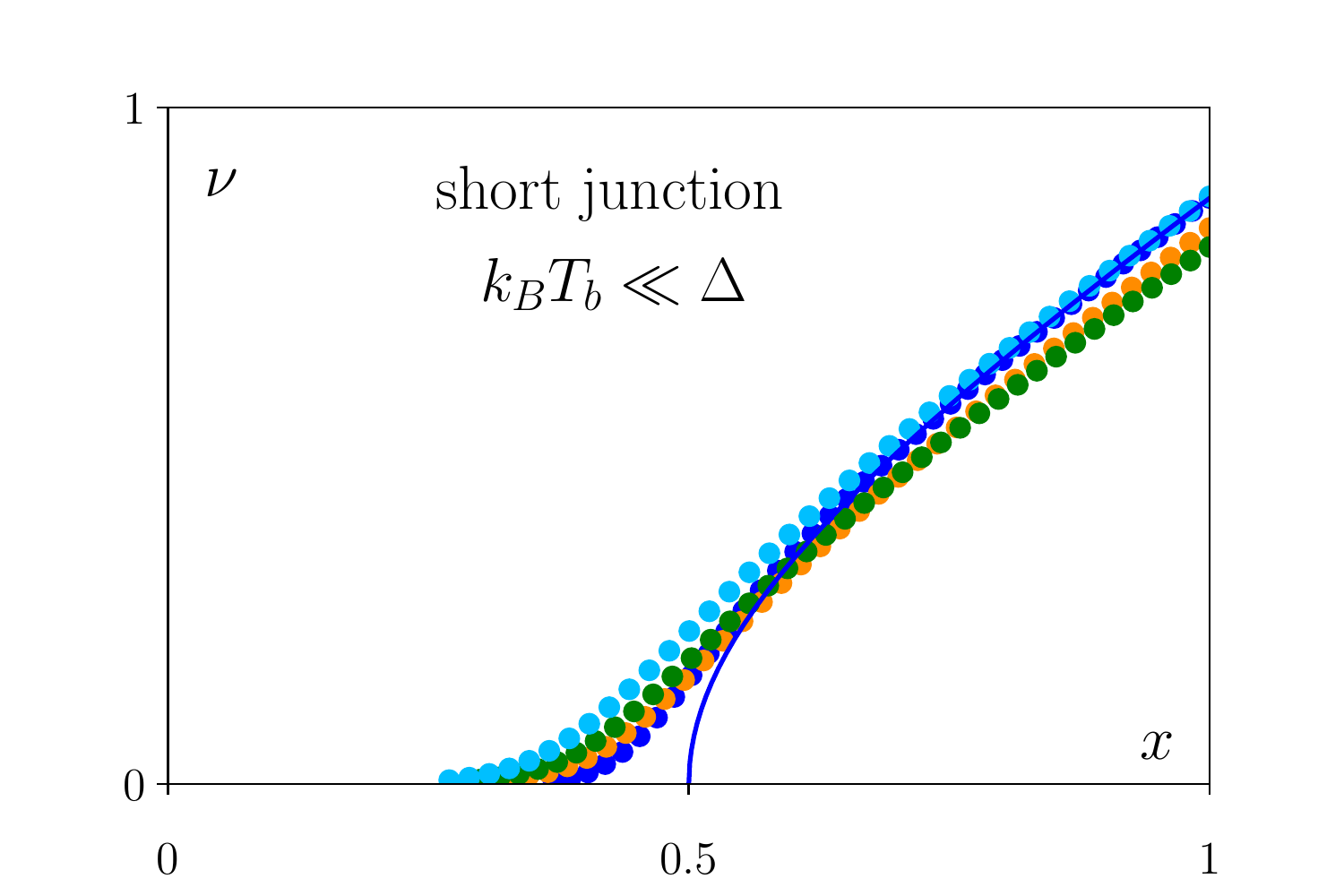}
		\label{fig:poisoningPaths:shortIVFPLowTemp}
	\end{subfigure}
	\hspace{-0.9cm}%
	\begin{subfigure}{\columnwidth}
		\vspace{-0.5cm}%
		\includegraphics[width = \columnwidth]{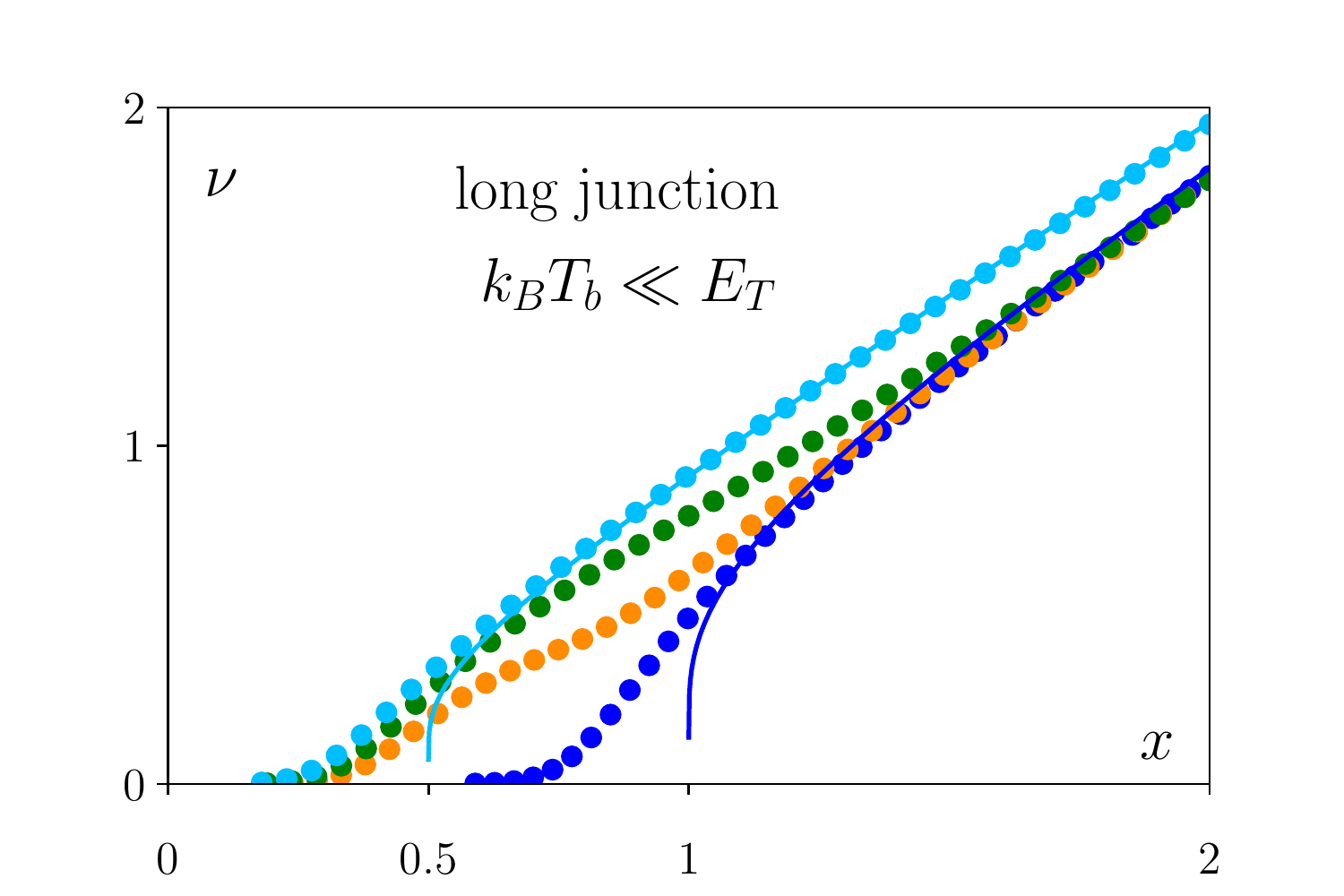}
		\label{fig:poisoningPaths:longIVFPLowTemp}
	\end{subfigure}
	\caption{Current voltage characteristics for the short (left) and long (right) junction limit for high (upper) and low (lower) PT $ T_b $ obtained by numerically solving the Fokker-Planck Eqs.~\eqref{eq:model:fokkerPlanck:FPE} (dotted lines) at junction temperature $ k_B T = 0.03\xi $. For large and small PRs the curves approaches the analytical solutions from Tab.~\ref{tab:poisoningEffect:solutions} (solid lines) for large bias currents.}
	\label{fig:fokkerPlanckSolutions}
\end{figure*}

\begin{figure*}
	\begin{subfigure}{\columnwidth}
		\includegraphics[width = \columnwidth]{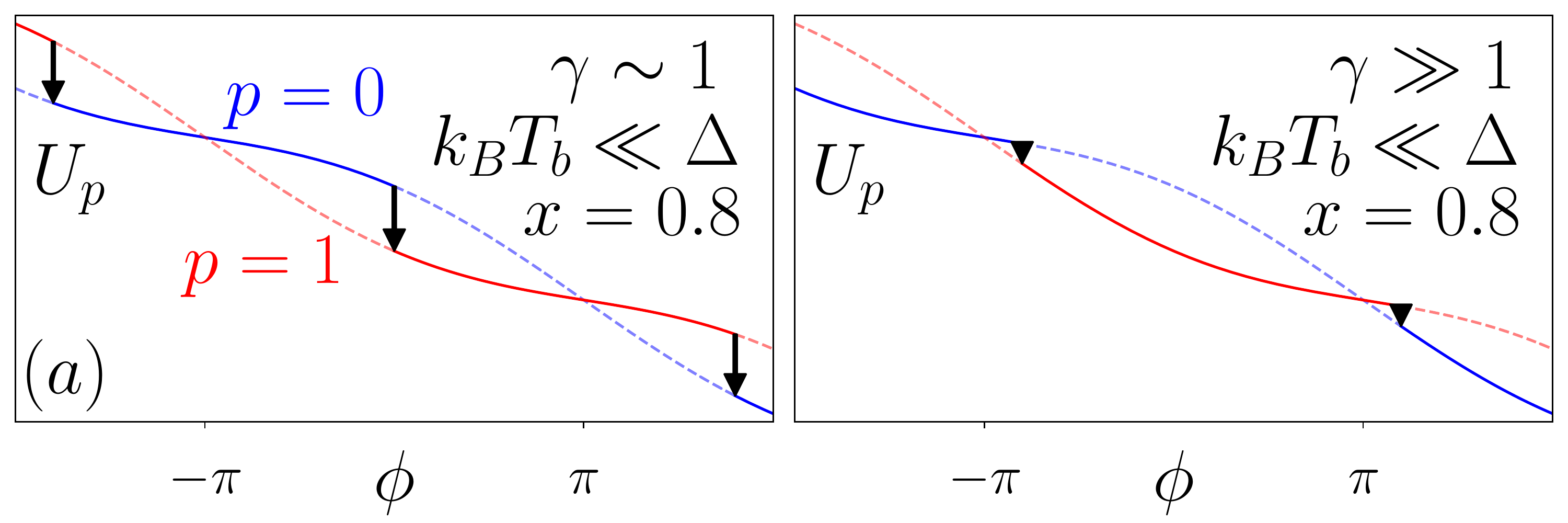}
	\end{subfigure}
	\begin{subfigure}{\columnwidth}
		\includegraphics[width = \columnwidth]{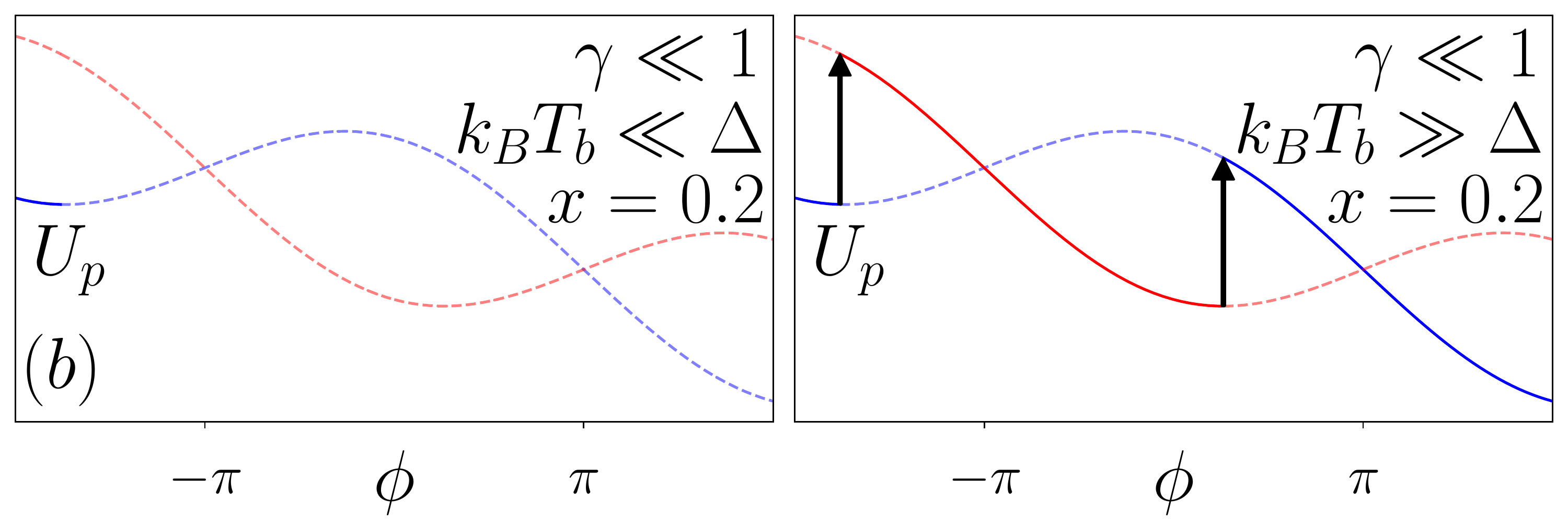}
	\end{subfigure}
	\begin{subfigure}{\columnwidth}
		\includegraphics[width = \columnwidth]{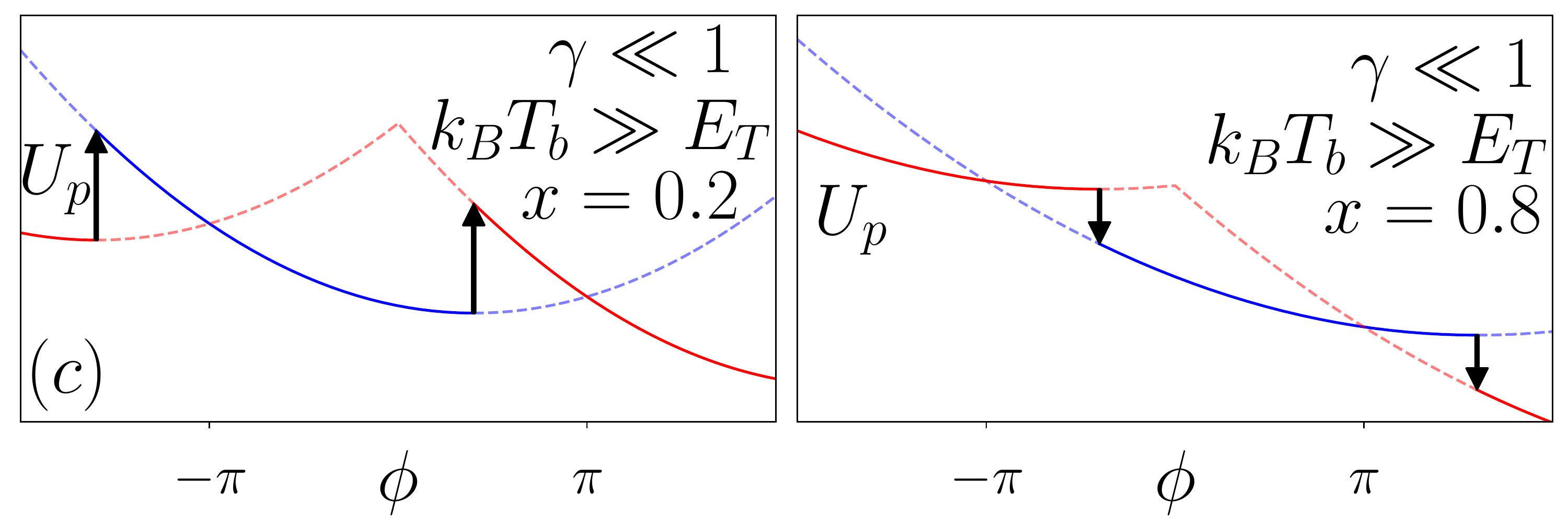}
	\end{subfigure}
	\begin{subfigure}{\columnwidth}
		\includegraphics[width = \columnwidth]{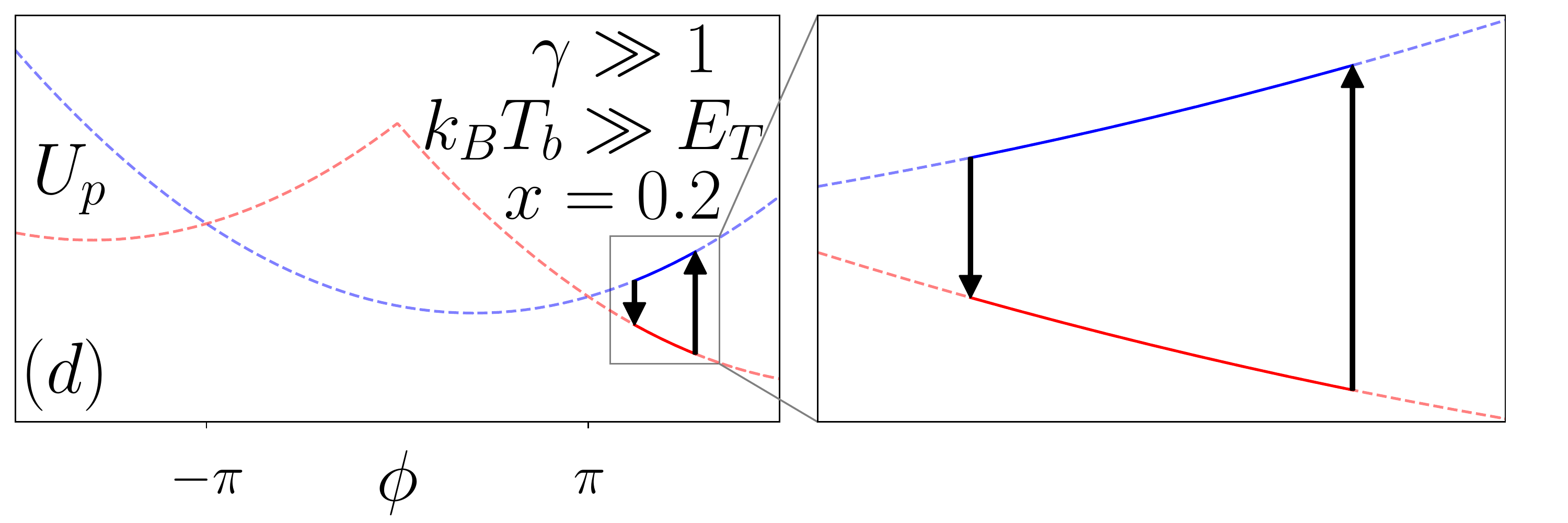}
	\end{subfigure}
	\caption{Example paths through which a junction can evolve for short (top) and long (bottom) junctions. The two colors correspond to the two different parities $ p = 0,1 $ of the junction where solid (dashed) lines indicate the occupied (unoccupied) state of the junction. When a poisoning event occurs, the junction changes its parity and therefore the branch in the above plots. In the case of $ k_B T_b \ll \xi $ predominantly poisoning events, which relax the junction to the potential of lower energy can occur whereas both relaxation and excitation events are possible for $ T_b \gg \xi $. For $ \gamma \ll 1 $ and bias currents below the critical current ((b) and (c)) the junction evolves to a potential minimum before a poisoning event takes place.}
	\label{fig:poisoningPaths}
\end{figure*}

In the case of small and large PRs $ \Gamma $ compared to the inverse of the intrinsic time scale $\tau_J$ of the junction and large bias currents $I$ the current voltage characteristics (Fig.~\ref{fig:fokkerPlanckSolutions}, dotted lines) approaches the corresponding analytical solutions of Tab.~\ref{tab:poisoningEffect:solutions} obtained by treating the poisoning via an effective potential $ U_{\textrm{eff}} $ (full lines). In the case of high PTs (Fig.~\ref{fig:fokkerPlanckSolutions}, upper panels) intermediate PRs interpolate between the two limits of vanishing and large PRs. For low PTs such a simple interpolation does not occur. In the short junction limit intermediate PRs lower the voltage $V$ that develops across the junction for high bias currents. A possible explanation of this effect is that the time scale between two poisoning events is comparable to the time scale over which the junction advances $ \phi $ by $ 2\pi $. Therefore, on average, approximately one poisoning event will occur between the two crossing points of the two potentials $ U_0 $ and $ U_1 $ (Fig.~\ref{fig:poisoningPaths}a, left). Since the slope of the lower potential (Fig.~\ref{fig:poisoningPaths}a, red line) is higher close to the crossing point $ \phi = -\pi $ and the poisoning event does not always occur right after the phase $ \phi $ has advanced beyond $ -\pi $ the junction is more likely to stay in the potential with a smaller slope. The smaller slope according to Eq.~\eqref{eq:model:fokkerPlanck:dgl} results in $ \phi $ advancing slower and, hence, in a lower voltage developing across the junction.

On the other hand, if the PR is large (Fig.~\ref{fig:poisoningPaths}a, right), the relaxation to the lower potential due to a poisoning event will occur shortly after the phase $ \phi $ has advanced beyond $ -\pi $ because these poisoning events are frequent compared to $\tau_J$. Therefore, the junction will spend more time in the potential with a large slope which results in a larger voltage drop.

In the long junction limit with low PT (Fig.~\ref{fig:fokkerPlanckSolutions}, bottom right panel) intermediate PRs ($ \gamma \sim 1 $) both extrapolate between the two analytical limits but also show the lowering of the voltage drop across the junction for high bias currents compared with the analytical result similar to the short junction limit.

\section{Zero temperature limit}
\label{sec:ciritcalCurrents}

The analysis of Sec.~\ref{sec:poisoningEffect} assumed a finite junction temperature $ T $. However, recent experiments explored junctions at temperatures far below the superconducting gap $ \Delta $ and Thouless energy $ E_T $\cite{Oostinga2013, Wiedenmann2015}. To gain insights into the dynamics of such junctions we turn to the zero temperature limit in this section for which we employ the second approach described in \ref{subsec:timeDependentParity}.

\begin{figure*}
	\begin{subfigure}{\columnwidth}
		\includegraphics[width = \columnwidth]{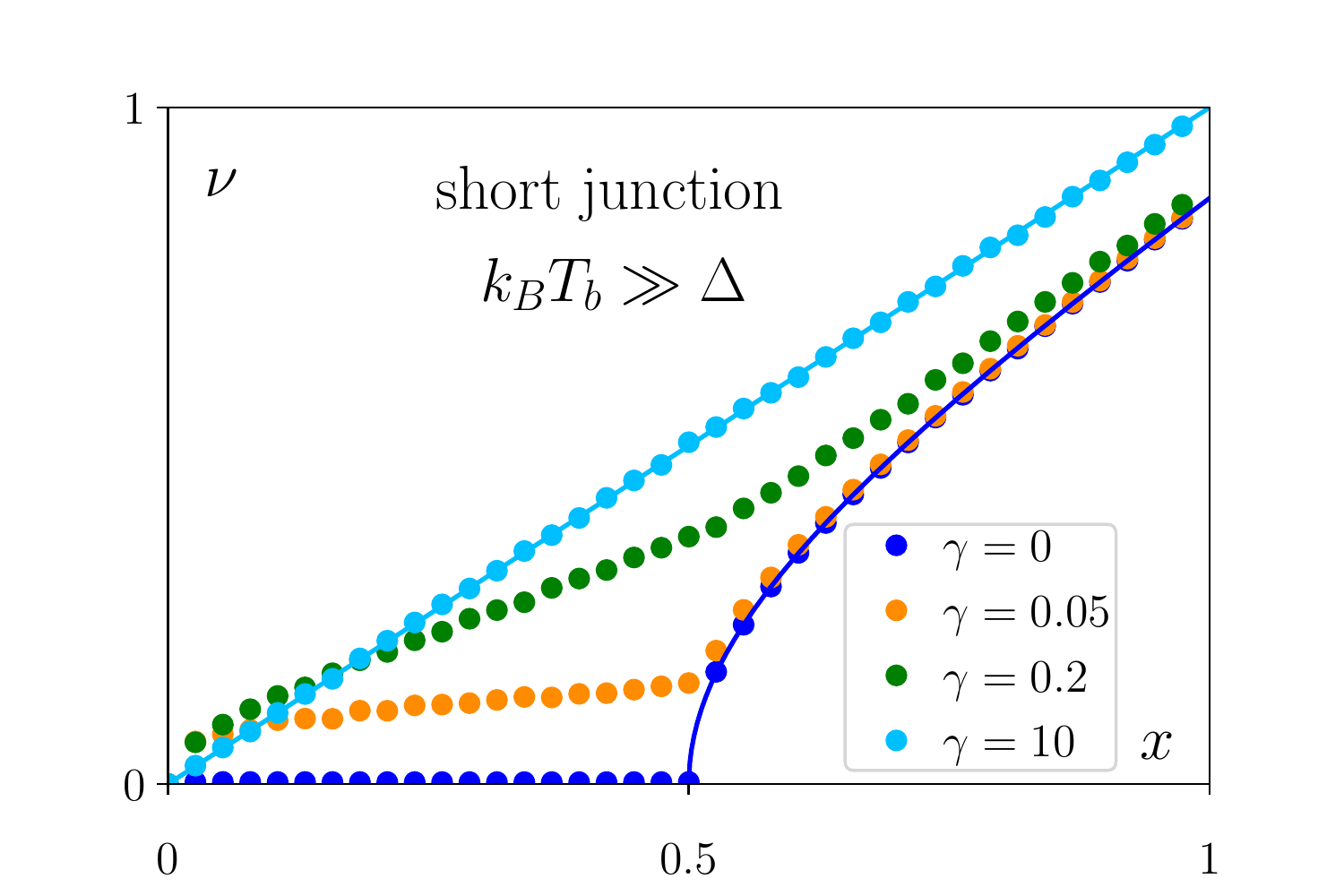}
		\label{fig:poisoningPaths:shortIVHighTemp}
	\end{subfigure}
	\hspace{-0.9cm}%
	\begin{subfigure}{\columnwidth}
		\includegraphics[width = \columnwidth]{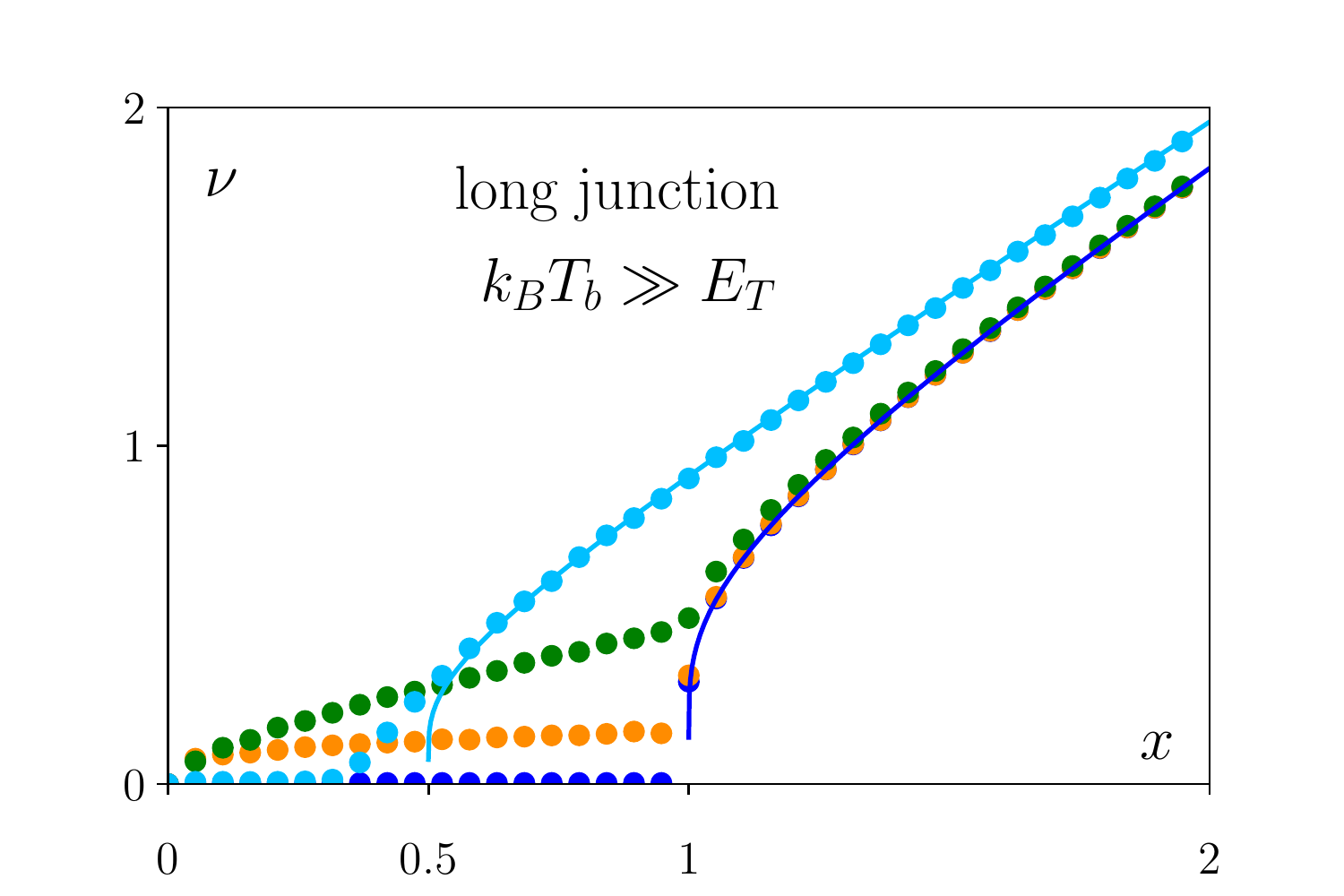}
		\label{fig:poisoningPaths:longIVHighTemp}
	\end{subfigure}
	\begin{subfigure}{\columnwidth}
		\vspace{-0.5cm}%
		\includegraphics[width = \columnwidth]{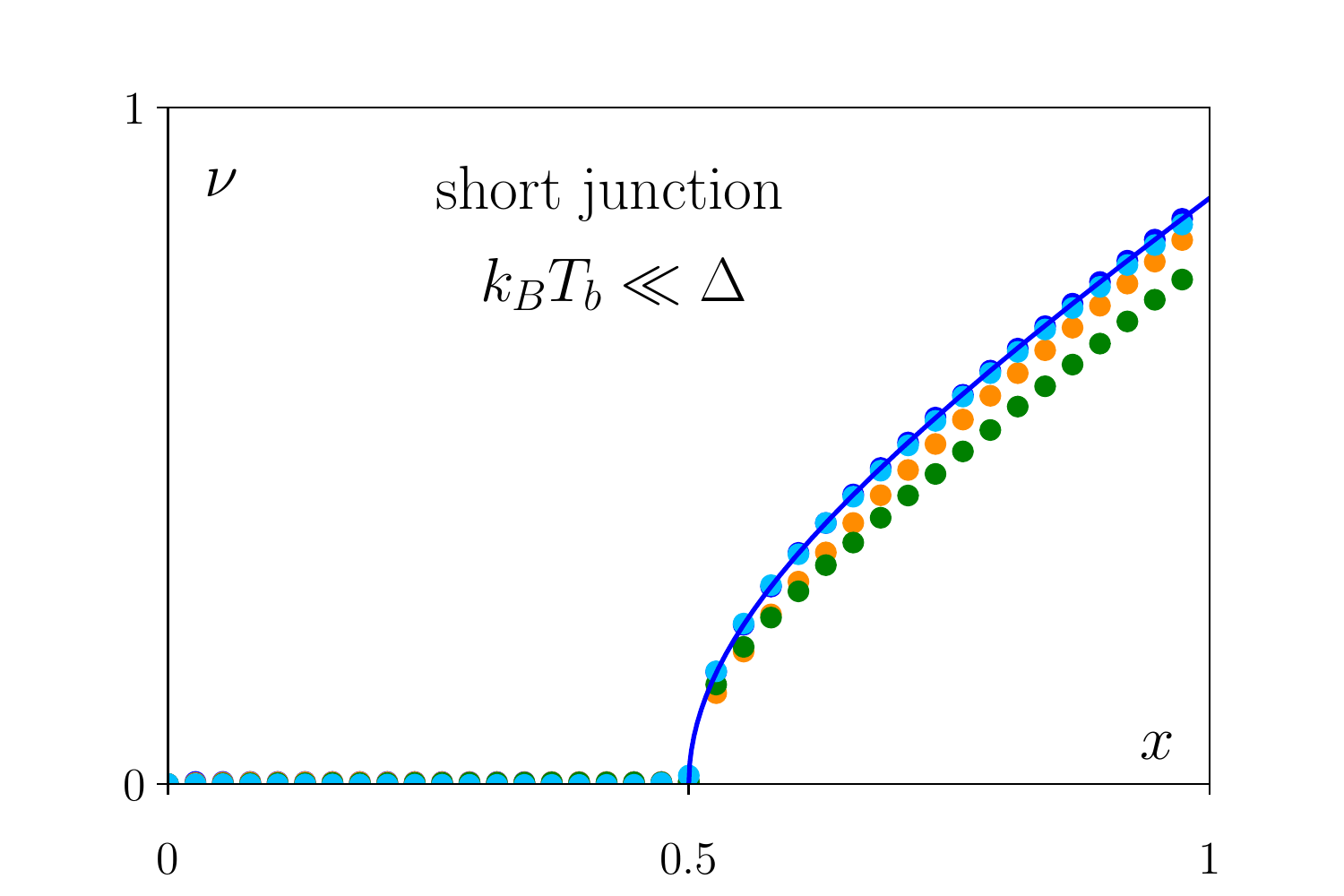}
		\label{fig:poisoningPaths:shortIVLowTemp}
	\end{subfigure}
	\hspace{-0.9cm}%
	\begin{subfigure}{\columnwidth}
		\vspace{-0.5cm}%
		\includegraphics[width = \columnwidth]{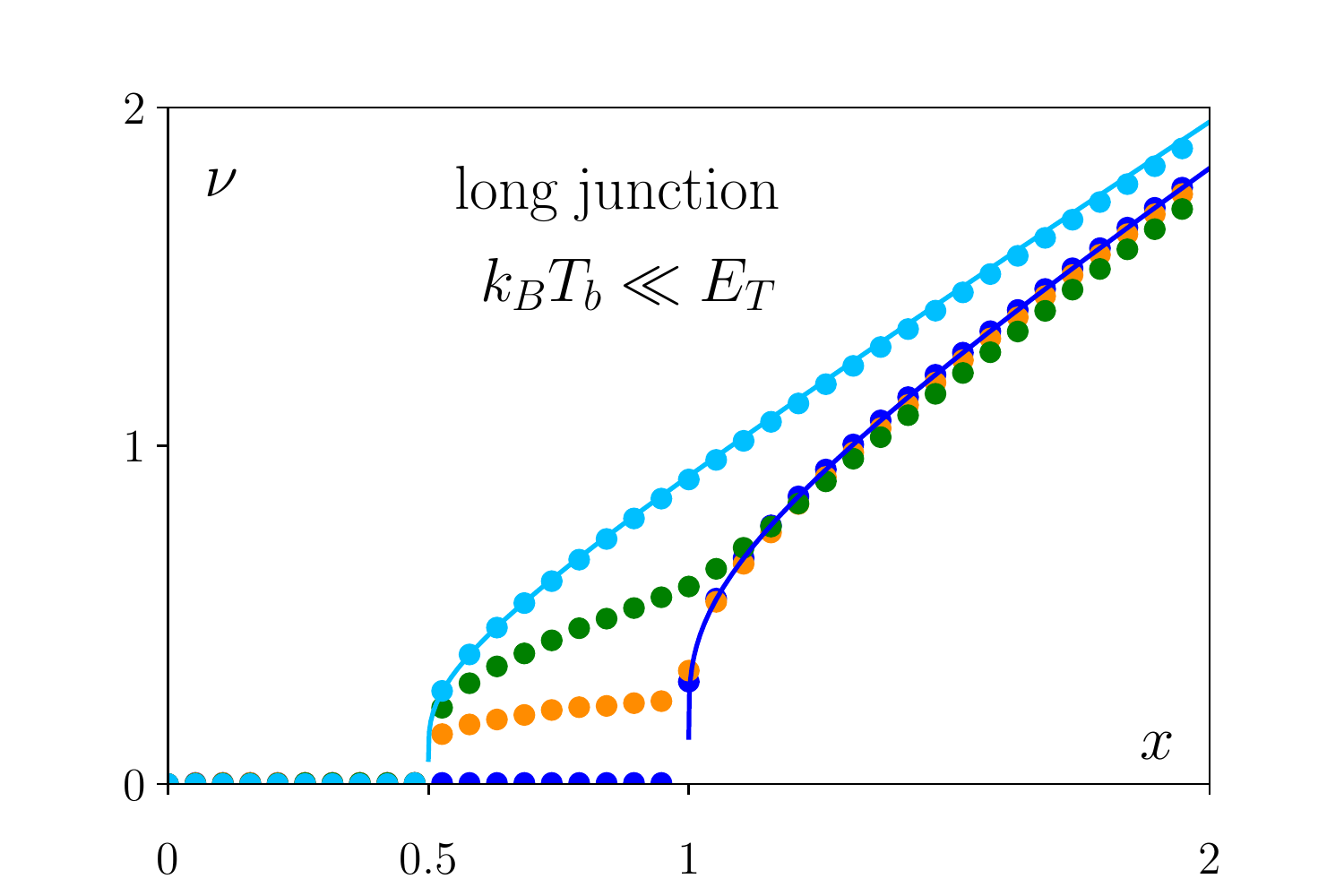}
		\label{fig:poisoningPaths:longIVLowTemp}
	\end{subfigure}
	\caption{Current voltage characteristic for the short (left) and long (right) junction limit for high (upper) and low (lower) PT obtained by numerically solving the differential equation~\eqref{eq:model:RSJDGL} (dotted lines) for a time dependent parity $ p(t) $ as described in \ref{subsec:timeDependentParity}. For large and small PRs the characteristic approaches the analytical solutions from Tab.~\ref{tab:poisoningEffect:solutions} (solid lines).}
	\label{fig:directSolutions}
\end{figure*}

In all cases (Fig.~\ref{fig:directSolutions}) the current voltage relation above the critical current behaves like the solutions of the Fokker-Planck equations.
Furthermore, the numerical results also follow the analytical solutions of Tab.~\ref{tab:poisoningEffect:solutions} for low bias currents in the case of no and large PRs.
In the following we will therefore specifically focus on the small bias current regime at zero temperature, but keep the PT as a finite parameter.

\subsection{Short junction}

Upon including a finite PR in the short junction limit with high PT (Fig.~\ref{fig:directSolutions}, upper left panel) a finite voltage drop develops for all bias currents below the initial critical current of $ x = 1/2 $ even for small PRs and the critical current therefore vanishes. For large PRs the current voltage characteristics follows the linear trend $ \nu = x $ as the junction becomes resistive, an effect which was already described in Ref.~\citenum{Lee2014}. On the other hand, the critical current in the case of low PT is unchanged by poisoning regardless of the strength of the PR. This difference can be explained by the fact that for finite bias currents the potential minima of one potential lies at the same phase as a negative slope of the other potential (Fig.~\ref{fig:poisoningPaths}b). If the system is in the energetically more favorable potential (\ref{fig:poisoningPaths}b, solid blue line for $ \phi \approx -2\pi $) the phase can only be advanced further and hence generate a finite voltage if the system gets excited into the energetically higher potential (red line). Such an excitation is only possible, if the poisoning electrons have a high temperature $ T_b $. A similar picture holds for the long junction limit to be discussed next.

\subsection{Long junction}

In the case of high PT in the long junction limit (Fig.~\ref{fig:directSolutions}, upper right panel) including a finite PR again results in a small but finite voltage drop across the junction for bias currents below the critical current in the absence of poisoning ($ x = 1 $) similar to the short junction limit. Increasing the PR further then suppresses the developing voltage for bias currents below half of the critical current without poisoning ($ x = 1/2 $). In contrast, the low PT case (Fig.~\ref{fig:directSolutions}, lower right) only ever develops a finite voltage for bias currents between $ x = 1/2 $ and $ x = 1 $. This can be explained again by considering the washboard potentials for the long junction limit (Fig.~\ref{fig:poisoningPaths}c). The system starts in the first energetically lower energy minimum (solid red line). To advance the phase beyond that point, the system must undergo an excitation (Fig.~\ref{fig:poisoningPaths}c, left) [relaxation (Fig.~\ref{fig:poisoningPaths}c, right)] in the case of $ x < 1/2 $ [ $ x > 1/2 $ ] due to a poisoning event. Since excitation events are only possible in the presence of high PT, only this case will develop a finite voltage for the bias currents $ x < 1/2 $. Because the phase can advance via relaxation processes for bias currents $ x > 1/2 $, a finite voltage will develop for both low and high PT even for small PRs.

The voltage drop across the junction in the long junction limit with high PT only approaches zero if the PR is much faster than the intrinsic time scale of the junction, which is explained by the phase relaxing to a position, where the two potentials have exactly opposite slopes (Fig.~\ref{fig:poisoningPaths}d). Because for large PRs the parity is flipped before the phase can evolve long distances due to the equation~\eqref{eq:model:RSJDGL} and the slopes of the two potentials only differ by a sign, the phase will rapidly move back and forth resulting in a voltage noise, but no finite voltage drop.

The change of the critical current from $ I_c $ ($ x = 1 $) to $ I_c / 2 $ ($ x = 1 / 2 $) in the long junction limit when increasing the PR has been suggested \cite{Beenakker2013} as a possible indicator for the topology of a Josephson junction assuming low PT. We find that this result can be extended to the high PT case. Furthermore distinguishing between a topological and non topological junction can only be successful, if the PR is much slower than the intrinsic time scale of the junction, because a small but finite voltage develops across the junction for $ 1 / 2 < x < 1 $ even for small poisoning.

\section{Poisoning Rates}
\label{sec:poisoningRates}

The above results show different behavior of the junctions depending on the PRs compared to the intrinsic time scale of the junction. However, it is also possible to extract the exact PR out of voltage measurements.

\subsection{Phase trapping}

\begin{figure}
\includegraphics[width = \columnwidth]{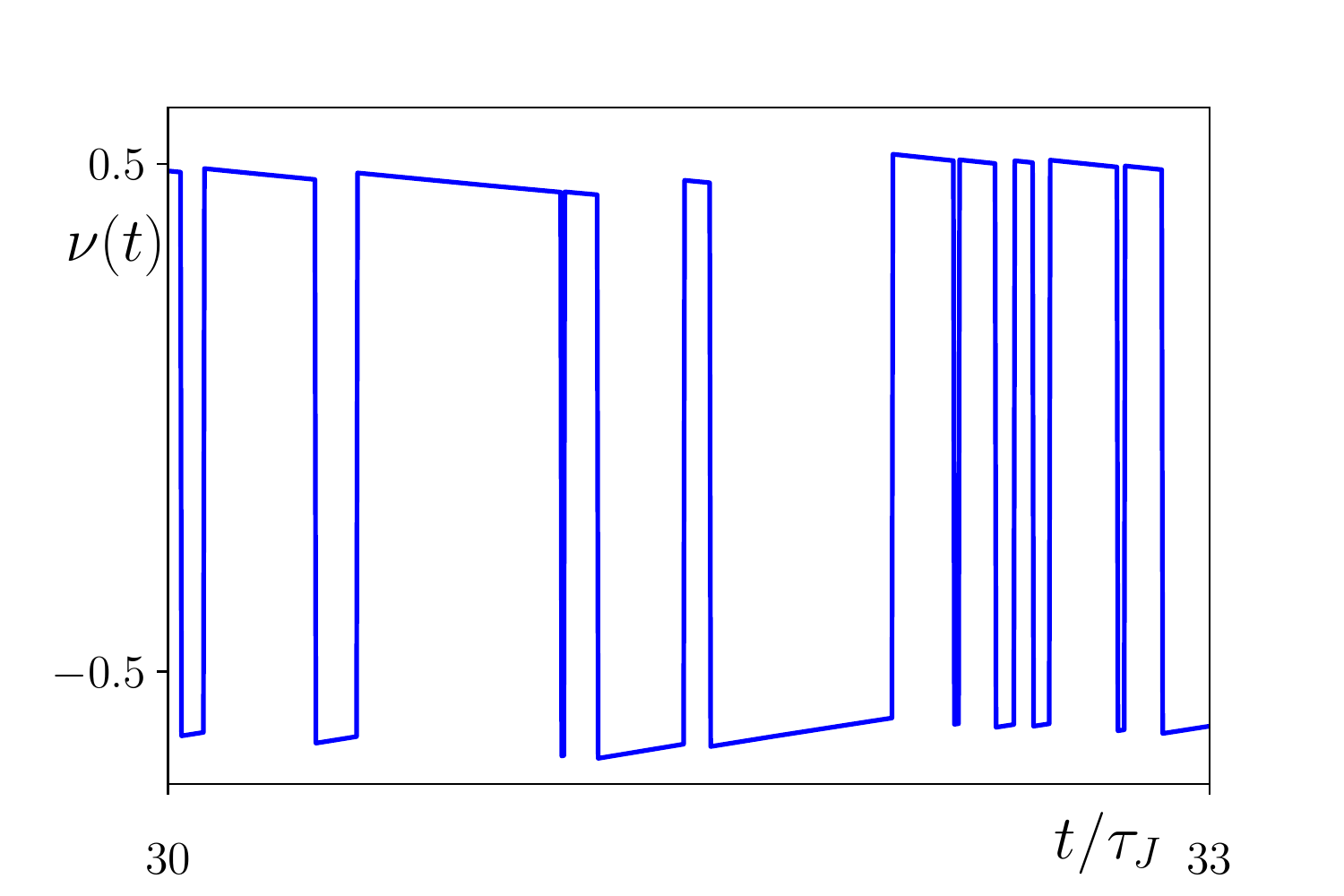}
\caption{Example of a time resolved voltage signal of the long junction limit with a large PR ($ \gamma = 10 $) and high PT for a current bias $ x = 0.2 $. The frequency of the voltage jumps is $ \gamma / 2 $.}
\label{fig:longJunctionHighPoisoningHighTempReadout}
\end{figure}

The first method to obtain the PR through a voltage measurement is only possible in long junctions due to their currents $ I_p(\phi) $ not being opposite for the two parities $ I_0(\phi) \neq - I_1(\phi) $.
If a long junction is in the large PR limit with high PT the PR can be obtained by performing a time resolved voltage measurement for bias currents $ x < 1/2 $. As explained in Sec.~\ref{sec:ciritcalCurrents} in this regime the phase $ \phi $ will relax to a position where the slopes of the two potentials $ U_p $ are exactly opposite. From there the system will frequently jump between the two potentials resulting in a phase which randomly moves back and forth but does not change its position when averaged over long time periods (Fig.~\ref{fig:poisoningPaths}d). This movement of the phase translates to a voltage signal which frequently jumps between $ V = \pm E_T eR/(2\hbar) $ ($ \nu = \pm 1/2 $) (Fig.~\ref{fig:longJunctionHighPoisoningHighTempReadout}). Since each voltage jump corresponds to a change of parity the rate of poisoning events $ \Gamma / 2 $ can be calculated by counting the number of jumps in a fixed time period.

\subsection{Phase diffusion}

\begin{figure}
\includegraphics[width = \columnwidth]{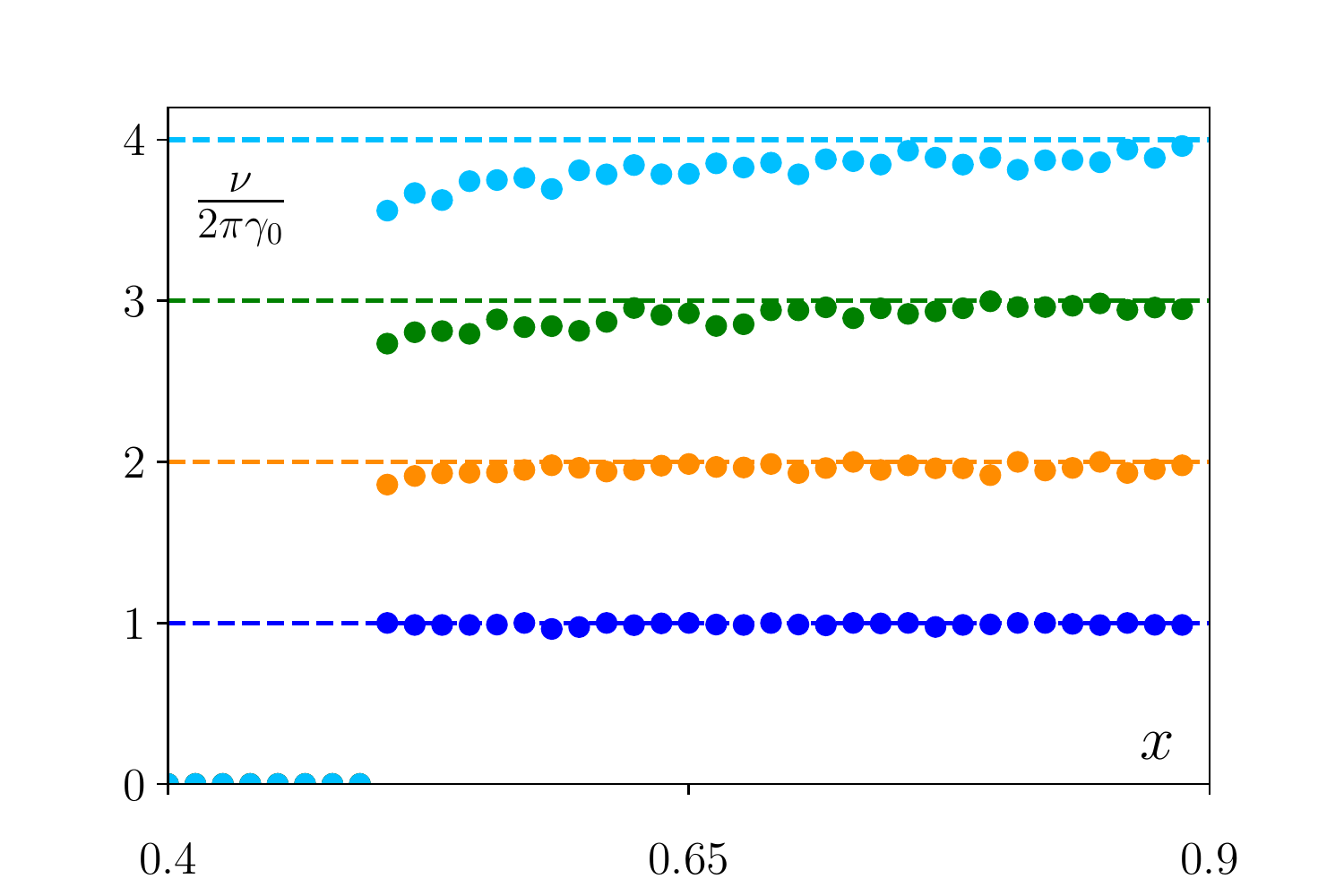}
\caption{Current voltage characteristic in the long junction regime for different small PRs ($ \gamma / \gamma_0 = 1, 2, 3, 4 $ with $ \gamma_0 = 10^{-3} $ from bottom to top) with low PT obtained by numerically solving the differential equation~\eqref{eq:model:RSJDGL} (dotted lines) for a time dependent parity $ p(t) $ as described in \ref{subsec:timeDependentParity}. The dashed lines are the expected voltage drops of $ \nu = 2\pi \gamma $.}
\label{fig:longJunctionHighPoisoningLowTempReadout}
\end{figure}

The above method to measure the PR is only possible in the large PR regime with high PT. However, in the case of small PRs there exists another method to obtain the PR if the junction is in the long junction regime or in the short junction with high PT regime. As an example we consider the low PT case in the long junction regime (the other two regimes are analogous). In this case the voltage developing across the junction follows the analytical solutions
given in Tab.~\ref{tab:poisoningEffect:solutions} for $ x > 1 $. In the range of intermediate bias currents $ 1 / 2 < x < 1 $ a small but finite voltage develops across the junction. For small PRs we can assume that the phase will relax to a potential minimum before another poisoning event occurs (Fig.~\ref{fig:poisoningPaths}c). After each poisoning event, the phase will relax to the next potential minimum. The distance between two minima of potentials of different parities is always $ 2\pi $ and most notably independent of the applied bias current. Each poisoning event is therefore associated with the phase advancing by $ 2\pi $ resulting in a small voltage pulse. These pulses can be counted in a time resolved voltage signal in a fixed time period to obtain the PR. Alternatively by time averaging the voltage signal a finite voltage of $ V = \hbar 2\pi \Gamma / (2e) $ ($ \nu = 2\pi\gamma $) will develop across the junction (Fig.~\ref{fig:longJunctionHighPoisoningLowTempReadout}) from which the PR can be obtained.

\subsection{Tunable poisoning rate}

\begin{figure}
\includegraphics[width = \columnwidth]{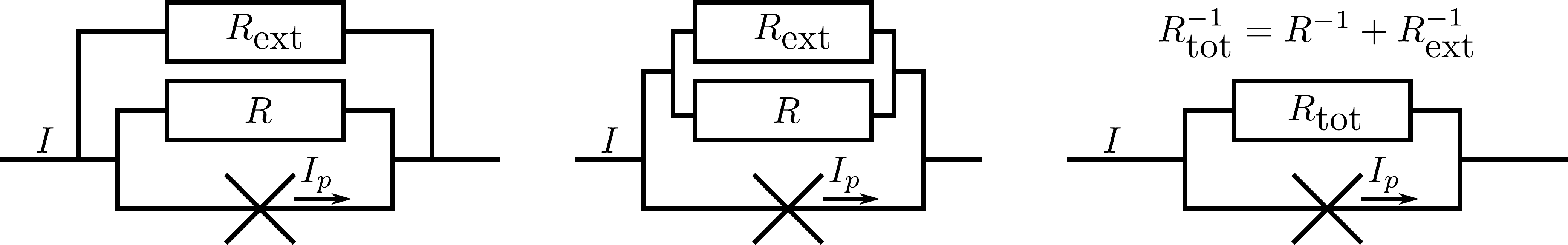}
\caption{Three equivalent circuits of a physical Josephson junction modeled by the RSJ model with an additional external resistance $ R_{\textrm{ext}} $. The rightmost circuit shows how this extended circuit is again an RSJ model with an adapted resistance $ R_{\textrm{tot}} $.}
\label{fig:RSJRExt}
\end{figure}

The only cases of the long junction limit in which the PR can not be obtained directly are the large PR with low PT and the intermediate PR regime. In the second scenario, however, we can connect a resistance $ R_{\textrm{ext}} $ in parallel to the junction modeled via the RSJ model (Fig.~\ref{fig:RSJRExt}, left). This circuit is equivalent with an RSJ model, where the intrinsic resistance $ R $ is replaced with two resistances $ R $ and $ R_{\textrm{ext}} $ in parallel (Fig.~\ref{fig:RSJRExt}, middle). Since resistances in parallel can be described with a single resistance $ R_{\textrm{tot}} $ with $ R_{\textrm{tot}}^{-1} = R^{-1} + R_{\textrm{ext}}^{-1} $ this is again equivalent to another RSJ model with an adapted resistance $ R_{\textrm{tot}} $ (Fig.~\ref{fig:RSJRExt}, right). By tuning the external resistance $ R_{\textrm{ext}} $ to be very small, the total resistance $ R_{\textrm{tot}} $ is also going to be small, which translates to a long intrinsic timescale of the Josephson junction $ \tau_J \propto R_{\textrm{tot}}^{-1} $.

With this, the case of small PR regime ($ \gamma \ll 1 $) as well as the cases in the intermediate PR regime ($ \gamma \approx 1 $) can be tuned to the large PR regime ($ \gamma \gg 1 $) by increasing the intrinsic time scale of the junction while the rate of poisoning events $ \Gamma $ stays constant. For example intrinsic quasiparticle poisoning events are predicted to occur on the time scale $ \tau_{qp} \sim \mu $s where typical time scales of Josephson junctions are on the order of $ \tau_J \sim 10 $ ps resulting in $ \gamma \sim 10^{-5} $. When changing the resistance $ R_\textrm{tot} $ the normalized PRs $ \gamma $ change according to $ R_\textrm{tot}/R_\textrm{tot}' = \gamma/\gamma' $. In order to tune the rate $ \gamma $ from the intrinsic case of $ \gamma \sim 10^{-5} $ to $ \gamma \sim 1 $ the resistance $ R_\textrm{tot} $ needs to decrease by 5 orders of magnitude. Since $ R_\textrm{tot} \ll R $ we can approximate $ R_\textrm{ext} \approx R_\textrm{tot} $ so that for typical resistances of $ R \sim 100 \, \Omega $ external resistances $ R_\textrm{ext} $ on the order of $ 1 \, $m$ \Omega $ are needed to achieve the intermediate poisoning regime $ \gamma \sim 1 $. Smaller resistances of the order of $ R_\textrm{ext} \sim 0.1 $~m$ \Omega $ are needed to achieve the high poisoning regime $ \gamma \sim 10 $.
Another way to tune the effective poisoning rate $ \gamma $ is to directly tune the coupling $ \Gamma $ to the reservoir from which the poisoning particles come from \cite{Frombach18}.

\section{Junction comprising two helical edge states with a constriction}
\label{sec:extendedJunction}

\begin{figure}
\includegraphics[width = 0.49\columnwidth]{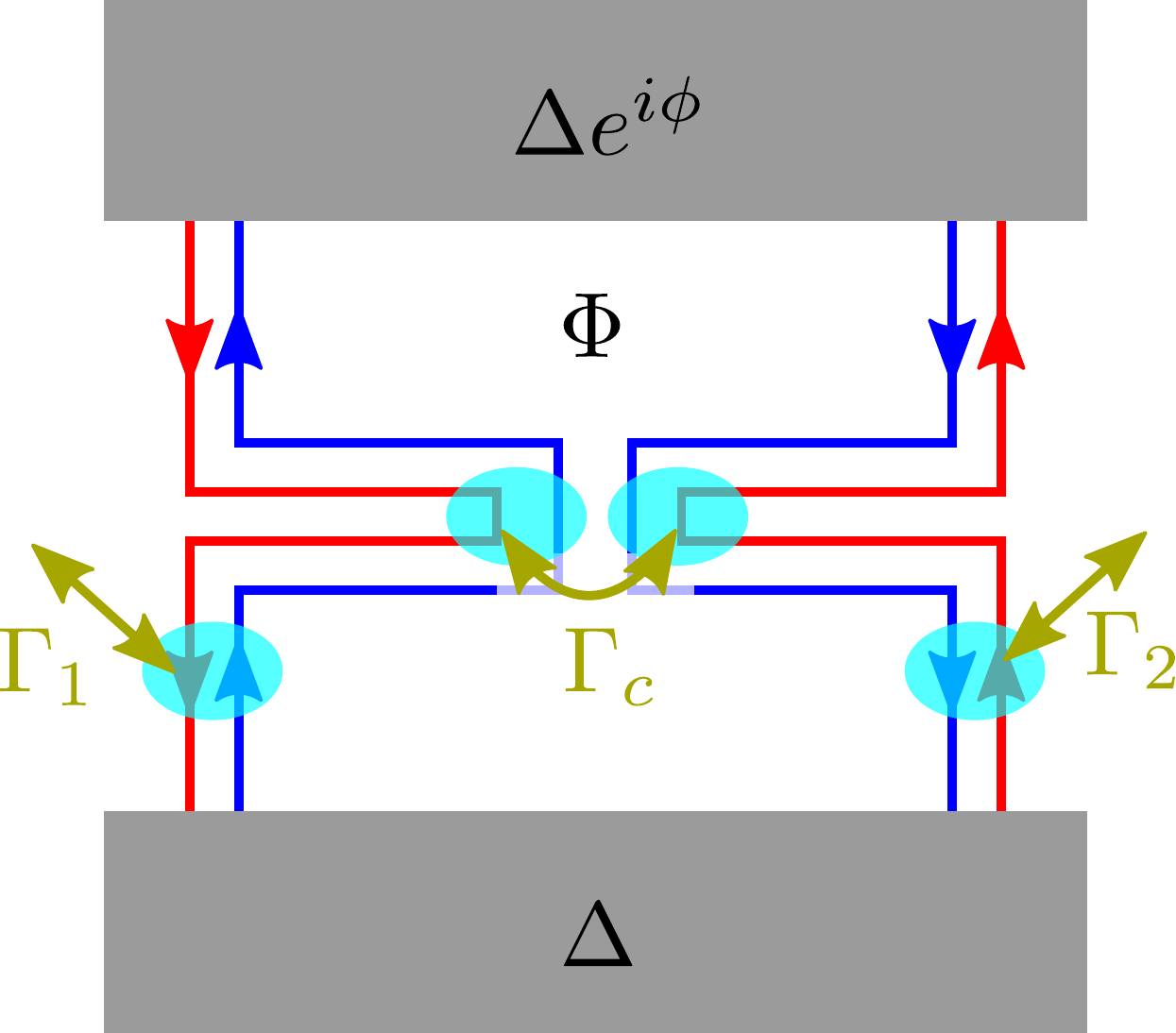}
\caption{
Setup of an extended Josephson junction, where the junction is mediated by two sets of quantum spin Hall edge states (blue and red).
A flux $ \Phi $ is threaded between the superconductors.
Quasiparticle poisoning events (dark yellow) can occur in both edge states independent of each other ($ \Gamma_{1,2} $).
Furthermore, in the middle restriction electrons can tunnel from one edge to the other ($ \Gamma_c $) which changes the parity of each edge but not the overall parity of the junction.
}
\label{fig:extendedJunction:setup}
\end{figure}

Up to now we have only considered a topological Josephson junction mediated by a single pair of quantum spin Hall edge states.
In an extended junction, where two edge states of both sample edges mediate the Josephson junction, signatures of the underlying topology have been found in the short junction limit even in the presence of quasiparticle poisoning \cite{Lee2014}.
We extend this model by also including a constriction (Fig.~\ref{fig:extendedJunction:setup}) in the edge states \cite{Strunz2019} mediating the Josephson junction to allow for electrons to tunnel from one edge to the other.
Such a tunneling event changes the parity of each edge but not of the overall junction. Furthermore quasiparticle poisoning events can occur independently in each edge with rates $ \Gamma_1 $ and $ \Gamma_2 $ which change the parity of one edge and therefore of the overall junction.
The parity dependent energy phase relation of the overall junction is therefore given by
\footnote{Here, we introduce an additional factor of $ 1/2 $ into the definition of the total energy of the junction to better compare the results with the ones obtained in the previous sections.
This effectively rescales $ \xi = E_T $ and therefore $ I_c $ by the same factor.}
\begin{equation}
\label{eq:extended:energy}
E_{p_1, p_2}(\phi) = \frac{E_{p_1}(\phi) + E_{p_2}(\phi + \Phi)}{2}
\end{equation}
where $ E_{p_i}(\phi) $ is the current phase relation of the junction edge $ i = 1, 2 $ and $ \Phi $ an additional flux threaded between the superconductors
\footnote{Here, $ \Phi = \pi \tilde{\Phi} / \Phi_0 $ with $ \Phi_0 = h/(2e) $ being the flux quantum and $ \tilde{\Phi} $ being the physical flux threaded between the superconductors.}.
The above mentioned quasiparticle PR $ \Gamma_1 $ ($ \Gamma_2 $) changes $ p_1 \rightarrow p_1 + 1 \mod 2 $ ($ p_2 \rightarrow p_2 + 1 \mod 2 $) and leaves $ p_2 $ ($ p_1 $) unchanged.
The tunneling of electrons from one edge to the other changes both parities $ p_1 \rightarrow p_1 + 1 \mod 2 $ and $ p_2 \rightarrow p_2 + 1 \mod 2 $ which leaves $ p_1 + p_2 \mod 2 $ unchanged.

\subsection{Spectral broadening}

Generally, the energy of the helical edge states of both edges will be broadened for instance through thermal broadening\footnote{
Other sources of broadening could include voltage fluctuations and the poisoning events in each individual edge resulting in a finite life time and therefore finite line width of the corresponding eigenstates of the junction.
While not every broadening effect turns the spectral density into a Lorentz-Cauchy distribution, the effects to be discussed are independent of the specific shape of the spectral density.
}
so that the density of states of the overall junction (see App.~\ref{app:spectralBroadening})
\begin{equation}
\label{eq:extended:broadening}
\rho(E) = \delta(E - E_{p_1, p_2}) \rightarrow \frac{1}{\pi} \frac{\frac{D}{2}}{(E-E_{p_1, p_2})^2 + \left( \frac{D}{2} \right)^2}
\end{equation}
changes to a Lorentz-Cauchy distribution.
The broadening $ D $ will generally depend on the specific sample and will be chosen in the following calculations as $ d = D/E_T = 0.1 $ to illustrate its effects.
The tunneling events then occur at a rate of (see App.~\ref{app:spectralBroadening})
\begin{equation}
W_c = \frac{\Gamma_c}{d} \frac{\left( \frac{d}{2} \right)^2}{\left( \frac{E_{p'_1, p'_2}-E_{p_1, p_2}}{E_T} \right)^2 + \left( \frac{d}{2} \right)^2}
\end{equation}
where $ \Gamma_c $ is a constant tunneling rate and $ p'_{i} = p_{i} + 1 \mod 2 $.

\subsection{Two critical currents}

In the long junction limit (Fig.~\ref{fig:extendedJunction:IV}) the current voltage characteristics are $ 2\pi $ periodic in the threaded flux $ \Phi $ due to poisoning similar to the short junction case \cite{Lee2014}.
By looking at the case for $ \Phi = 0 $ one can make out two 'critical currents' in this setup.
A first around $ x = 1/2 $ at which a finite voltage develops across the junction.
However, a second 'critical current' exists around $ x = 1 $ at which point the voltage shows a stronger increase with increasing bias currents.
This forming of two critical currents is independent of the tunneling between the two edges and also present in the absence of tunneling events \cite{Lee2014} as energy conservation during tunneling events restricts the parameter regime in which tunneling events can have an influence on the voltage developing across the junction.

The parity of the junction can be described by the vector $ \vec{p} = (p_1, p_2) $ and can be in one of four states.
In the case $ \Phi = 0 $ the overall even states $ (0, 0) $ and $ (1, 1) $ feature a $ 4\pi $ periodic current phase relation with a critical current of $ x = 1 $ whereas the overall odd states $ (0, 1) $ $ (1, 0) $ have a $ 2\pi $ periodic current phase relation and a critical current of $ x = 1/2 $ \cite{Crepin14}.
In the latter two states the junction is resistive for bias currents $ 1/2 < x < 1 $ and a finite voltage
\begin{equation}
	\nu_o = \left[ 
	\ln \left( \frac{x + \frac{1}{2}}{x - \frac{1}{2}} \right)
	\right]^{-1}
\end{equation}
develops in the absence of poisoning, which can be obtained by solving the corresponding RSJ model analytically.
Because the junction can freely move between all parity configurations via poisoning events the voltage drop across the junction is given by
\begin{equation}
\nu = \tau_o \nu_o
\end{equation}
where $ \tau_o $ is the fraction of the time the junction spends in an overall odd state.

\subsection{Voltage peak due to tunneling events}

While the tunneling events between the two edges leaves the current phase relation unchanged for $ x \not\approx 1/2 $, a voltage peak centered around $ x = 1/2 $ develops due to the tunneling events, which is independent of the flux threaded between the two superconductors.
In the case of $ \Phi = 0 $, however, the already finite voltage drop at $ x \gtrsim 1/2 $ overshadows this effect.

\begin{figure}
\includegraphics[width = \columnwidth]{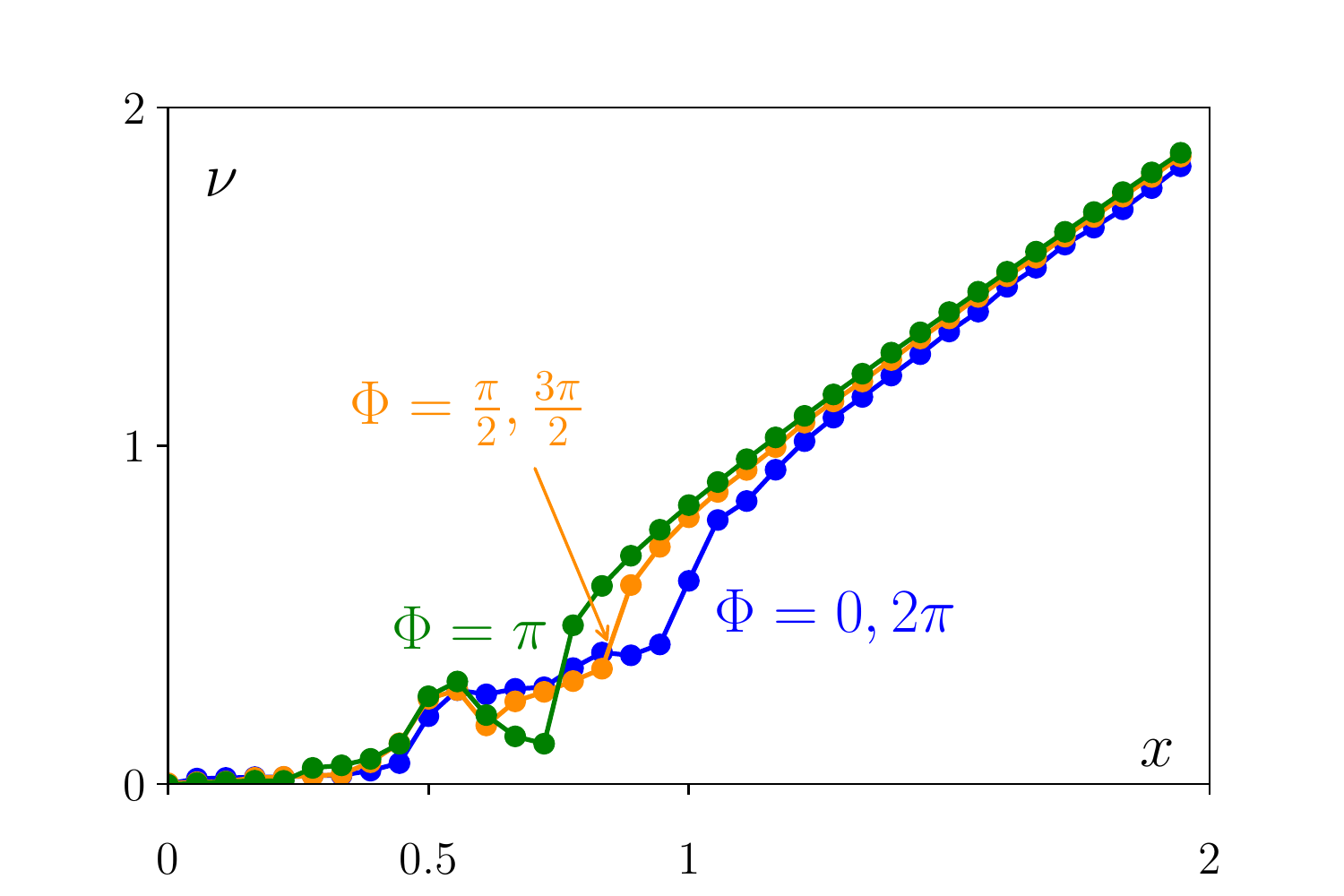}
\includegraphics[width = \columnwidth]{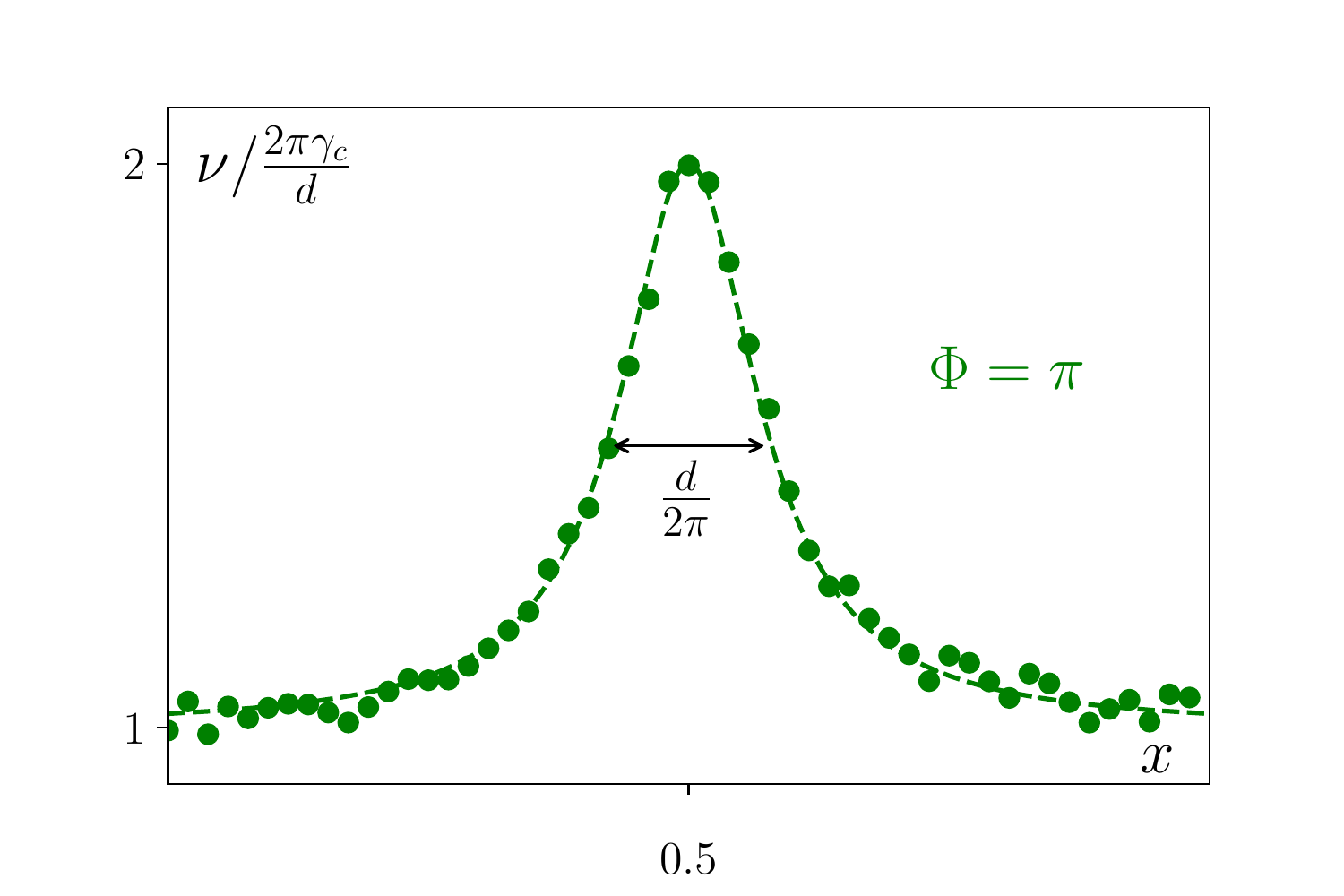}
\caption{
Current voltage characteristic for the extended junction (Fig.~\ref{fig:extendedJunction:setup}) in the long junction limit and spectral broadening of the form of Eq.~\eqref{eq:extended:broadening}.
Tunneling events between the two edges occur with the rate $ \gamma_c = \Gamma_c \tau_J = 0.1 $ ($ \gamma_c = 10^{-5} $) while the PR is equal in both edges and equal to $ \gamma_1 = \gamma_2 = 10^{-2} $ ($ 10^{-4} $) at the top (bottom).
Different fluxes $ \Phi $ are threaded between the two superconductors.
The dashed line (bottom) is a Lorentz curve given by the analytical prediction Eq.~\eqref{eq:extended:noPoisoningLorentzian}.
}
\label{fig:extendedJunction:IV}
\end{figure}

For small tunneling rates $ \gamma_c $ (Fig.~\ref{fig:extendedJunction:IV}, bottom), i.e. when the time scale between two tunneling events is much larger than the intrinsic time scale of the junction
\footnote{In addition, the time it takes for an electron-hole pair in an Andreev bound state to make a full round trip $ 2L/v_F $ should be much smaller than the timescale on which tunneling events happen $ (\Gamma_c/d)^{-1} $, which is true in the parameter regime discussed in this paper.},
the existence of the voltage peak can be explained by the parity dependent washboard potentials $ U_{p_1 p_2}(\phi) $.
Tunneling events between the two edges can only occur at phases, where the potentials $ U_{00}(\phi) $ and $ U_{11}(\phi) $ or the potentials $ U_{01}(\phi) $ and $ U_{10}(\phi) $ cross, because only at these phases does a tunneling event conserve the energy.
Since the junction with an overall odd parity is equivalent to a junction with overall even parity with a shifted flux $ \Phi \rightarrow \Phi + 2\pi $ it is sufficient to only consider the first case.
The potentials $ U_{00}(\phi) $ and $ U_{11}(\phi) $ cross at the phases $ \phi_{c1} = \pi -\Phi/2 $ and $ \phi_{c2} = 3\pi -\Phi/2 $.
If the junction starts in the state, where both edges have even parity, the overall junction will evolve according to the washboard potential $ U_{00}(\phi) $.
If the tunneling rate $ \gamma_c $ is small, the junction will relax to the first minimum of the potential $ U_{00}(\phi) $ located at $ \phi_{\textrm{min}} = 2\pi x - \Phi/2 $ before a potential tunneling event can take place.
In the case of $ x = 1/2 $, this minimum coincides with the crossing point of the two potentials, so that a tunneling event can take place and the phase can relax to the next minimum of the potential $ U_{11}(\phi) $ located at $ \phi = 2\pi x - \Phi/2 + 2\pi $, i.e. exactly $ 2\pi $ further.
From this point on the entire process can repeat, resulting in a diffusion of the phase down the washboard potentials.
Since each tunneling event results in an advancement of the phase $ \phi $ by $ 2\pi $ and the tunneling events occur with a frequency of $ \Gamma_c/d $ the voltage drop across the junction for $ x = 1/2 $ is given by $ \nu = 2\pi \gamma_c / d $.
If the bias current is close to but not exactly equal to $ x = 1/2 $ there exists a phase difference of $ \pi(2x - 1) $ between the crossing point $ \phi_{c1} $ and the minimum $ \phi_{\textrm{min}} $ which results in an energy difference of $ \approx E_T \pi(2x - 1) $ between the two potentials $ U_{00}(\phi) $ and $ U_{11}(\phi) $ at the phase $ \phi_\textrm{min} $.
In an isolated system, the conservation of energy (expressed through a delta function in the tunneling rates) would prohibit such a tunneling event.
In real samples however broadening effects of the delta function mentioned in Eq.~\eqref{eq:extended:broadening} allow for these events to occur, albeit with a lower probability.
The resulting voltage drop across the junction
\begin{equation}
\nu_{\textrm{tun}} = 2\pi \frac{\gamma_c}{d} \frac{\left( \frac{d}{4\pi} \right)^2}{\left( x - \frac{1}{2} \right)^2 + \left( \frac{d}{4\pi} \right)^2}
\end{equation}
therefore also takes the shape of a Lorentzian around $ x = 1/2 $ with a maximum of $ \nu_{\textrm{tun}} = 2\pi \gamma_c / d $ at $ x = 1/2 $ and a full width at half maximum (FWHM) of $ d/(2\pi) $.

To measure this voltage peak the threaded flux must differ from $ \Phi = 0 $ in order to distinguish the peak from the finite voltage developing above the critical current.
In the case of $ \Phi = \pi $ a voltage drop of $ \nu_{\textrm{pois}} = 2\pi\gamma_1 $ develops for small PRs $ \gamma_1 = \gamma_2 $ for bias currents $ 1/4 < x < 3/4 $ through the same mechanisms described in sec.~\ref{sec:ciritcalCurrents}.
Because the two effects are independent of each other the resulting voltages simply add up so that the overall voltage developing across the junction in the vicinity of $ x = 1/2 $ will therefore be given by
\begin{equation}
\label{eq:extended:noPoisoningLorentzian}
\nu = \nu_{\textrm{tun}} + \nu_{\textrm{pois}}
= 2\pi \frac{\gamma_c}{d} \frac{\left( \frac{d}{4\pi} \right)^2}{\left( x - \frac{1}{2} \right)^2 + \left( \frac{d}{4\pi} \right)^2}
+ 2\pi\gamma_1 .
\end{equation}

\section{Conclusion}
\label{sec:conclusion}
In this work, we have studied the current voltage characteristics of fractional Josephson junctions, concentrating on the effect of parity changing quasiparticle poisoning using the RSJ model. We treated poisoning as originating from a particle reservoir with its own temperature that can be high or low compared to the intrinsic energy scale of the Josephson junction. We modeled the poisoning source by an effective equilibrium distribution of quasiparticles at a given temperature $T_b$. We stress that the assumption of an equilibrium distribution is not essential for our purpose as long as the poisoning quasiparticles are at an energy $\sim k_BT_b$. Using the Fokker-Planck equation at finite (but small) junction temperatures, we extended earlier results in the short junction regime \cite{Lee2014} to the long-junction regime. We thereby included the lowest energy state of each parity sector for the tilted washboard potential. This is valid for temperatures below the Thouless energy and fast relaxation times (or low PT) within each parity sector. At zero junction temperature, we developed another method to calculate the current voltage characteristics using a time-dependent parity and integrated the RSJ model equation. 

We found that the distinct critical currents in the long-junction limit of large and small poisoning rates (PRs) predicted in \cite{Beenakker2013} can be found also at high poisoning temperatures (PT). We have devised ways to directly measure the PR in the long junction regime, either employing the average voltage signal in the regime of small PRs or via the time frequency of voltage pulses in the large PR limit.
In contrast to the short junction regime, these measurements are possible in long junctions both in the low and high PT case.
The knowledge of such PRs are essential for predicting the stability of topological qubits made from Majorana fermions in these systems. 

In a Josephson junction made from two pairs of helical edge states with a magnetic flux threading the junction, a constriction between the two edges is shown to provide another tool to characterize a fractional Josephson junction. We find a voltage peak at exactly half the critical current of a long Josephson junction with a height that is proportional to the tunneling rate through the constriction. The width of this peak is bounded (from below) by the sum of the PRs from each edge. This voltage peak can be distinguished from the background voltage across the Josephson junction using the magnetic flux through the junction. 

In future works, the applied dc-current could be supplemented with a ac-component to study the Shapiro effect as well as non-equilibrium populations of the Andreev bound states in the junction. 

\section{Acknowledments}
We thank Fernando Dom\'{i}nguez for very useful feedback on the manuscript, and Wolfgang Belzig and Markus Etzkorn for enlightening discussions.
We acknowledge financial support from the Braunschweig-Hannover cluster of excellence "Quantum Frontiers - Light and Matter at the Quantum Frontier" (EXC 2123), the Lower Saxony PhD-programme “Contacts in Nanosystems”, the Research Training Group GrK1952/1 “Metrology for Complex Nanosystems” and the Braunschweig International Graduate School of Metrology B-IGSM.
\begin{appendix}
	
\section{Spectral broadening and tunneling rate through the constriction}
\label{app:spectralBroadening}

In the setup of Sec.~\ref{sec:extendedJunction} a spectral broadening $D$ of the total energy of the junction was introduced. Here, we show how this spectral broadening 
can be derived from the spectral broadening of each edge (assumed to be a Lorentz-Cauchy distribution) using the method of Fermi's golden rule. The resulting effect will be a Fermi's golden rule rate with a Lorentz-Cauchy distribution for the total energy of the junction (see also Eq.~\eqref{eq:extended:broadening}) where the width $D$ is given by the sum of all individual spectral broadenings in each edge and parity $p$.

The tunneling rate for quasiparticles through the constriction is given by
\begin{equation}
	W_c = \frac{2\pi}{\hbar} \sum_{if} \abs{\braket{f|H_T|i}}^2 \delta[\varepsilon_i - \varepsilon_f]
\end{equation}
where $ H_T $ is a tunneling Hamiltonian that transfers particles between the Andreev bound states comprised of the two pairs of helical edge states. The initial and final states are both product states of the states of each individual edge
\begin{equation}
	\ket{i} = \ket{p_1} \otimes \ket{p_2},
	\qquad
	\ket{f} = \ket{p'_1} \otimes \ket{p'_2}
\end{equation}
where $ p_{1/2} $ indicate the parity of the left / right edge before the tunneling. After the tunneling event the parity of both edges change to $ p'_{1,2} = p_{1,2} + 1 \mod 2 $. Because the sum over all initial and final states runs over a continuum the sums can be converted into integrals
\begin{equation}
	\begin{aligned}
		W_c = \frac{2\pi}{\hbar} \int d\varepsilon_{p_1} \rho_{p_1} \, d\varepsilon_{p_2} \rho_{p_2} \, d\varepsilon_{p'_1} \rho_{p'_1} \, d\varepsilon_{p'_2} \rho_{p'_2} \\
		\abs{\braket{f|H_T|i}}^2 \delta[(\varepsilon_{p_1} + \varepsilon_{p_2})  - (\varepsilon_{p'_1} + \varepsilon_{p'_2})],
	\end{aligned}
\end{equation}
where
\begin{equation}
	\label{eq:app:spectralBroadening:LCDDef}
	\rho_{\alpha} = \frac{1}{\pi} \frac{\gamma_{\alpha}}{(\varepsilon_{\alpha} - E_{\alpha})^2 + \gamma_{\alpha}^2}
	=: L[\varepsilon_{\alpha} - E_{\alpha}, \gamma_{\alpha}]
\end{equation}
is the spectral density of the state $ \alpha = p_1, p_2, p'_1, p'_2 $ and has the form of a Lorentz Cauchy distribution with a full width at half maximum (FWHM) of $ 2\gamma_{\alpha} $ around the unperturbed energy $ E_{\alpha} $. Assuming that the energy dependence of the matrix elements can be neglected for the energies involved the problem reduces to solving the integral
\begin{equation}
	\begin{gathered}
		\int d\varepsilon_{p_1} d\varepsilon_{p_2} d\varepsilon_{p'_1} d\varepsilon_{p'_2}
		L[\varepsilon_{p_1} - E_{p_1}, \gamma_{p_1}] \\
		L[\varepsilon_{p_2} - E_{p_2}, \gamma_{p_2}]
		L[\varepsilon_{p'_1} - E_{p'_1}, \gamma_{p'_1}] \\
		\hspace{1cm}L[\varepsilon_{p'_2} - E_{p'_2}, \gamma_{p'_2}]
		\delta[(\varepsilon_{p_1} + \varepsilon_{p_2})  - (\varepsilon_{p'_1} + \varepsilon_{p'_2})].
	\end{gathered}
\end{equation}
Performing the first integral over $ \varepsilon_{p'_2} $ eliminates the delta function
\begin{equation}
	\label{eq:app:spectralBroadening:threeLCD}
	\begin{gathered}
		\int d\varepsilon_{p_1} d\varepsilon_{p_2} d\varepsilon_{p'_1}
		L[\varepsilon_{p_1} - E_{p_1}, \gamma_{p_1}] \\
		L[\varepsilon_{p_2} - E_{p_2}, \gamma_{p_2}]
		L[\varepsilon_{p'_1} - E_{p'_1}, \gamma_{p'_1}] \\
		\hspace{1cm}L[\varepsilon_{p_1} + \varepsilon_{p_2} - \varepsilon_{p'_1} - E_{p'_2}, \gamma_{p'_2}].
	\end{gathered}
\end{equation}
Using the relationship (for a proof, see App.~\ref{app:intOverTwoLCD})
\begin{equation}
	\int d\varepsilon L[\varepsilon + a, \gamma] L[\varepsilon + b, \delta]
	= L[a - b, \gamma + \delta]
\end{equation}
and $ L[\varepsilon, \gamma] = L[-\varepsilon, \gamma] $ we can integrate over $ \varepsilon_{p'_1} $ to reduce equation~\eqref{eq:app:spectralBroadening:threeLCD} to
\begin{equation}
	\begin{gathered}
		\int d\varepsilon_{p_1} d\varepsilon_{p_2}
		L[\varepsilon_{p_1} - E_{p_1}, \gamma_{p_1}]
		L[\varepsilon_{p_2} - E_{p_2}, \gamma_{p_2}] \\
		L[\varepsilon_{p_1} + \varepsilon_{p_2} - E_{p'_1} - E_{p'_2}, \gamma_{p'_1} + \gamma_{p'_2}].
	\end{gathered}
\end{equation}
Now the same relationships can be used to first integrate over $ \varepsilon_{p_2} $
\begin{equation}
	\begin{gathered}
		\int d\varepsilon_{p_1}
		L[\varepsilon_{p_1} - E_{p_1}, \gamma_{p_1}] \\
		L[-\varepsilon_{p_1} + E_{p'_1} + E_{p'_2} - E_{p_2}, \gamma_{p_2} + \gamma_{p'_1} + \gamma_{p'_2}]
	\end{gathered}
\end{equation}
and finally over $ \varepsilon_{p_1} $
\begin{equation}
	\begin{aligned}
		&L[- E_{p_1} + E_{p'_1} + E_{p'_2} - E_{p_2}, \gamma_{p_1} + \gamma_{p_2} + \gamma_{p'_1} + \gamma_{p'_2}] \\
		=& L[ (E_{p'_1} + E_{p'_2}) - (E_{p_1} + E_{p_2}), \gamma_{p_1} + \gamma_{p_2} + \gamma_{p'_1} + \gamma_{p'_2}] \\
		=& L[ E_{p'_1, p'_2} - E_{p_1, p_2}, \gamma_{p_1} + \gamma_{p_2} + \gamma_{p'_1} + \gamma_{p'_2}] \\
		=& \frac{1}{\pi} \frac{ \frac{D}{2} }{ (E_{p'_1, p'_2} - E_{p_1, p_2})^2 + \left( \frac{D}{2} \right)^2 },
	\end{aligned}
\end{equation}
where we have used the notation defined in Eq.~\eqref{eq:extended:energy} and introduced the total broadening $ D/2 = \gamma_{p_1} + \gamma_{p_2} + \gamma_{p'_1} + \gamma_{p'_2} $. The tunneling rate therefore takes the form
\begin{equation}
	W_c = \frac{2\pi}{\hbar} \abs{\braket{f|H_T|i}}^2 \frac{1}{\pi} \frac{ \frac{D}{2} }{ (E_{p'_1, p'_2} - E_{p_1, p_2})^2 + \left( \frac{D}{2} \right)^2 }
\end{equation}
which again takes the form of Fermi's golden rule with a spectral function
\begin{equation}
\rho \rightarrow \frac{1}{\pi} \frac{ \frac{D}{2} }{ (E_{p'_1, p'_2} - E_{p_1, p_2})^2 + \left( \frac{D}{2} \right)^2 }
\end{equation}
for the entire system.

Writing $ \abs{t_c} = \braket{f|H_T|i} $ and $ d = D / E_T $ we can further simplify the rate
\begin{equation}
\begin{aligned}
W_c 
&= \frac{2\pi}{\hbar} \abs{t_c}^2 \frac{1}{\pi} \frac{\frac{D}{2}}{(E_{p'_1, p'_2}-E_{p_1, p_2})^2 + \left( \frac{D}{2} \right)^2} \\
&= \frac{2\pi}{\hbar} \abs{t_c}^2 \frac{1}{\pi} \frac{1}{E_T} \frac{\frac{d}{2}}{\left( \frac{E_{p'_1, p'_2}-E_{p_1, p_2}}{E_T} \right)^2 + \left( \frac{d}{2} \right)^2} \\
&= \frac{2\pi}{\hbar} \abs{t_c}^2 \frac{1}{\pi} \frac{1}{E_T} \frac{2}{d} \frac{\left( \frac{d}{2} \right)^2}{\left( \frac{E_{p'_1, p'_2}-E_{p_1, p_2}}{E_T} \right)^2 + \left( \frac{d}{2} \right)^2} \\
&= \frac{\Gamma_c}{d} \frac{\left( \frac{d}{2} \right)^2}{\left( \frac{E_{p'_1, p'_2}-E_{p_1, p_2}}{E_T} \right)^2 + \left( \frac{d}{2} \right)^2}
\end{aligned}
\end{equation}
with
\begin{equation}
\Gamma_c = \frac{2\pi}{\hbar} \abs{t_c}^2 \frac{2}{\pi E_T}.
\end{equation}
The tunneling rate therefore also has a Lorentzian shape with a maximum of $ \Gamma_c / d $ and a FWHM of $ d $. We note that $D/\hbar$ is bounded (from below) by the quasiparticle poisoning rate (PR) $\Gamma_i$, $i=1,2$ of each individual helical edge channel for small coupling between the edges (small $\Gamma_c$).

\section{Integral over two Lorentz Cauchy Distributions}
\label{app:intOverTwoLCD}

In the derivation of App.~\ref{app:spectralBroadening} we used the relationship
\begin{equation}
	\int d\varepsilon L[\varepsilon + a, \gamma] L[\varepsilon + b, \delta]
	= L[a - b, \gamma + \delta]
\end{equation}
multiple times where $L[\cdot,\cdot]$ is the Lorentz Cauchy distribution defined in Eq.~\eqref{eq:app:spectralBroadening:LCDDef}. To show this, we rewrite the Lorentz Cauchy distributions like
\begin{equation}
	L[\varepsilon, \gamma] := \frac{1}{\pi} \frac{\gamma}{\varepsilon^2 + \gamma^2}
	= \frac{\gamma}{\pi} \frac{1}{(\varepsilon + i\gamma)}\frac{1}{(\varepsilon - i\gamma)}
\end{equation}
so that
\begin{equation}
	\label{eq:app:intOverTwoLCD:complexIntegral}
	\begin{aligned}
		\int &d\varepsilon L[\varepsilon + a, \gamma] L[\varepsilon + b, \delta] \\
		&= \frac{\gamma \delta}{\pi^2} \int d\varepsilon 
		\frac{1}{(\varepsilon + a + i\gamma)}\frac{1}{(\varepsilon + a - i\gamma)} \\
		&\hspace{3cm}\frac{1}{(\varepsilon + b + i\delta)}\frac{1}{(\varepsilon + b - i\delta)}.
	\end{aligned}
\end{equation}
Performing the integral in the complex plane we can close the contour in the upper complex half plane ($ \Im(\varepsilon) > 0 $). This contour encompasses two of the four poles located at $ \varepsilon = -a + i\gamma $ and $ \varepsilon = -b + i\delta $. The residue of the integrand at these two poles is
\begin{equation}
	\frac{1}{2i\gamma} \frac{1}{(b - a) + i(\gamma + \delta )} \frac{1}{(b - a) + i(\gamma - \delta)}
\end{equation}
and
\begin{equation}
	\begin{aligned}
		&\frac{1}{(a - b) + i(\delta + \gamma)} \frac{1}{(a - b) + i(\delta - \gamma)} \frac{1}{2i\delta} \\
		=& \frac{1}{2i\delta} \frac{1}{(b - a) - i(\delta + \gamma)} \frac{1}{(b - a) + i(\gamma - \delta)},
	\end{aligned}
\end{equation}
respectively. The sum of both residues yields
\begin{equation}
	\begin{aligned}
		&\frac{1}{2i\gamma} \frac{1}{(b - a) + i(\gamma + \delta )} \frac{1}{(b - a) + i(\gamma - \delta)} \\
		&+ \frac{1}{2i\delta} \frac{1}{(b - a) - i(\delta + \gamma)} \frac{1}{(b - a) + i(\gamma - \delta)} \\
		=& \frac{1}{2i} \frac{1}{(b - a) + i(\gamma - \delta)} \\
		&\left[ \frac{1}{\gamma} \frac{1}{(b - a) + i(\gamma + \delta )} + \frac{1}{\delta} \frac{1}{(b - a) - i(\delta + \gamma)} \right] \\
		=& \frac{1}{2i} \frac{1}{(b - a) + i(\gamma - \delta)} \frac{1}{\gamma \delta} \\
		& \frac{\delta [(b - a) - i(\delta + \gamma)] + \gamma [(b - a) + i(\gamma + \delta )]}{(b - a)^2 + (\gamma + \delta )^2} \\
		=& \frac{1}{2i} \frac{1}{(b - a) + i(\gamma - \delta)} \frac{1}{\gamma \delta} \\
		& \frac{(b - a) (\delta + \gamma) + i (\delta + \gamma)(\gamma - \delta)}{(b - a)^2 + (\gamma + \delta )^2} \\
		=& \frac{1}{2i} \frac{1}{(b - a) + i(\gamma - \delta)} \frac{1}{\gamma \delta}
		\frac{(\delta + \gamma) [(b - a) + i (\gamma - \delta)]}{(b - a)^2 + (\gamma + \delta )^2} \\
		=& \frac{1}{2i} \frac{1}{\gamma \delta} \frac{(\delta + \gamma)}{(b - a)^2 + (\gamma + \delta )^2} \\
	\end{aligned}
\end{equation}
so that the integral~\eqref{eq:app:intOverTwoLCD:complexIntegral} evaluates to
\begin{equation}
	\begin{aligned}
		\int &d\varepsilon L[\varepsilon + a, \gamma] L[\varepsilon + b, \delta] \\
		&= \frac{\gamma \delta}{\pi^2} 2\pi i \frac{1}{2i} \frac{1}{\gamma \delta} \frac{(\delta + \gamma)}{(b - a)^2 + (\gamma + \delta )^2} \\
		&= \frac{1}{\pi} \frac{(\delta + \gamma)}{(b - a)^2 + (\gamma + \delta )^2} \\
		&= L[b - a, \gamma + \delta] \\
		&= L[a - b, \gamma + \delta].
	\end{aligned}
\end{equation}
	
\end{appendix}
\newpage

\end{document}